\documentclass[reviewcopy]{elsarticle}

\usepackage[reviewcopy]{adndt}
\usepackage{longtable}

\usepackage{mathptmx}
\usepackage{amsmath}
\usepackage{amssymb}

\biboptions{square,sort&compress}
\bibpunct[]{[}{]}{,}{n}{}{;}
\newcommand{\al}{$\alpha$}

\newcommand{\raa}{($\alpha$,$\alpha$)}
\newcommand{\raX}{($\alpha$,$X$)}

\newcommand{\rag}{($\alpha$,$\gamma$)}
\newcommand{\ran}{($\alpha$,n)}

\newcommand{\rap}{($\alpha$,p)}

\newcommand{\rga}{($\gamma$,$\alpha$)}

\newcommand{\stot}{$\sigma_{\rm{reac}}$}
\newcommand{\sred}{$\sigma_{\rm{red}}$}
\newcommand{\ered}{$E_{\rm{red}}$}

\begin{document}

\begin{frontmatter}

\journal{Atomic Data and Nuclear Data Tables}

\title{Elastic alpha scattering experiments and the alpha-nucleus optical
  potential at low energies}

  \author[One,Two]{P. Mohr\corref{cor1}}
  \ead{E-mail: mohr@atomki.mta.hu}

  \author[One]{G.\ G.\ Kiss}

  \author[One]{Zs.\ F\"ul\"op} 

  \author[Three]{D.\ Galaviz}

  \author[One]{Gy.\ Gy\"urky}

  \author[One]{E.\ Somorjai}

  \cortext[cor1]{Corresponding author.}

  \address[One]{Institute of Nuclear Research (ATOMKI), H-4001 Debrecen, Hungary}

  \address[Two]{Diakonie-Klinikum, D-74523 Schw\"abisch Hall, Germany}

  \address[Three]{Centro de F\'isica Nuclear, University of Lisbon, P-1649-003
    Lisbon, Portugal}

\date{12.12.2012}

\begin{abstract}
High precision angular distribution data of \raa\
elastic scattering are presented for the nuclei
$^{89}$Y, $^{92}$Mo, $^{106,110,116}$Cd, $^{112,124}$Sn, and $^{144}$Sm at
energies around the Coulomb barrier. Such data with small experimental
uncertainties over the full angular range (20-170 degrees) are the
indispensable prerequisite for the extraction of local optical potentials and
for the determination of the total reaction cross section \stot .

A systematic fitting procedure was applied to the presented experimental
scattering data to obtain comprehensive local potential parameter sets
which are composed of a real folding potential and an imaginary potential of
Woods-Saxon surface type. The obtained potential parameters were used in
turn to construct a new systematic \al -nucleus potential with very few
parameters. Although this new potential cannot reproduce the angular
distributions with the same small deviations as the local potential, the new
potential is able to predict the total reaction cross sections for
all cases under study.
\end{abstract}

\end{frontmatter}



\newpage

\tableofcontents
\listofDtables
\listofDfigures
\vskip5pc

\section{Introduction}
\label{int}
The \al -nucleus potential is the key ingredient for the calculation of \al
-particle induced reaction cross sections. For intermediate mass and heavy
target nuclei the reaction cross sections are typically
calculated within the framework
of the statistical model where the cross section 
$\sigma$\raX\ of an \al -induced reaction
is given by the product of the compound formation cross section
$\sigma_{CF}$ and the decay branching $b_X$ into the $X$ channel. Usually, the
compound formation cross section $\sigma_{CF}$ is approximated by the total
reaction cross section \stot\ which depends on the transmission coefficient
$T_\alpha$ of the incoming \al -particle. The decay branching is given by $b_X
= T_X / \sum_i T_i$ where the summation has to be performed over all open
channels $i$ (including $T_\alpha$). This leads to the well-known
proportionality
\begin{equation}
\sigma(\alpha,X) \sim \frac{T_\alpha T_X}{\sum_i T_i} \quad \quad .
\label{eq:statmod}
\end{equation}
In many cases it turns out that the transmission into other channels is much
larger than the transmission into the \al\ channel. This holds in particular
for the neutron channel as soon as the energy exceeds the threshold of the
\ran\ reaction. In such cases we find $T_X \approx \sum_i T_i$, and the cross
section $\sigma$\raX\ in 
Eq.~(\ref{eq:statmod}) becomes proportional to $T_\alpha$, i.e.\ it is mainly
defined by the transmission coefficient into the \al\ channel and the
underlying \al -nucleus potential. 
The full formalism of the statistical model
is reviewed e.g.\ in \cite{Rau00,Rau11}.

It has been noticed that predictions of \al -induced reaction cross sections
at low energies
have large uncertainties depending on the chosen \al -nucleus potential. This
holds in particular for \rag\ capture reactions for targets with masses above
$A \approx 100$ \cite{Som98,Gyu06,Ozk07,Cat08,Yal09,Gyu10,Kiss11,Kiss11a} but
also for \ran\ reactions as e.g.\ seen recently in \cite{Sau11,Mohr11}. In
addition, 
the uncertainties of \rag\ cross sections translate into similar uncertainties
for the prediction of \rga\ cross sections and \rga\ reaction rates under the
stellar conditions of the astrophysical $p$- or $\gamma$-process
\cite{Mohr07,Kiss08, Rau09}. The stellar reaction rates of these photon-induced
\rga\ reactions are usually determined from the inverse \rag\ capture cross
sections using 
detailed balance because thermally excited states contribute significantly
to the stellar reaction rate of photon-induced reactions \cite{Mohr06,Rau11};
these thermally excited states are not accessible for \rga\ experiments in the
laboratory.

Despite significant effort over the last decades, there is still no global
potential available that is able to describe elastic scattering angular
distributions and cross sections of \al -induced reactions simultaneously with
high quality. Up to now, the four-parameter energy-independent potential by
McFadden and Satchler is still widely used \cite{McF66}. A many-parameter
potential has been developed by Avrigeanu and coworkers over the last decade
\cite{Avr10,Avr09}. A simple potential for higher energies is given by Kumar
{\it{et al.}}\ \cite{Kum06} which has turned out to be inadequate at very low
energies. A further global potential has been developed by Demetriou {\it{et
al.}}\ \cite{Dem02}, and a simple 6-parameter potential was optimized for
reaction cross sections at low energies by Rauscher \cite{Rau03}. An attempt
was made to find a common potential for \al\ decay and \al\ capture
\cite{Den09}. Very
recently, a further regional potential in the cadmium/tin/tellurium region has
been derived from elastic scattering data by the Notre Dame astrophysics group
and coworkers \cite{Pal08,Pal12}.

Elastic scattering is a standard method for the determination of \al -nucleus
optical potentials. However, at energies close to the Coulomb barrier
ambiguities are found for the derived potentials because of the dominating
Rutherford contribution to the scattering cross section. It is obvious that
experimental data have to be measured with high precision to reduce the
ambiguities as much as possible. Here we present a summary of all the data
that have been measured over the last 15 years at the cyclotron of ATOMKI,
Debrecen, together with a new and consistent analysis within the optical model.

The paper is organized as follows: In Sect.~\ref{sec:exp} we briefly present
our experimental set-up and the data analysis. Sect.~\ref{sec:opt} provides
some information on the theoretical analysis and the derived local parameters
of the optical potential. All experimental data are listed in the Tables at
the end and are available as supplementary content. 
A suggestion for a new \al -nucleus potential is given in the Appendix
(Sect.~\ref{sec:new}).
All energies are given as $E_{\rm{c.m.}}$ in the center-of-mass system
except explicitly noted.

\section{Experimental technique}
\label{sec:exp}

\subsection{Beam properties and targets}

All alpha elastic scattering experiments discussed in the present paper were
carried out at the cyclotron laboratory of ATOMKI, Debrecen. In this paper a
brief overview on the experimental technique is given. Further experimental
details can be found in the original papers \cite{moh97, ful01, kis06, gal05,
  kis08, kis09, kis11}, conference proceedings \cite{kis_omeg, ful_omeg} and
in a Ph.D. dissertation \cite{gal_phd}. The proton and neutron number, the
chemical form of the target material, isotopic enrichment of the target, the
energy of the first excited states of the target nuclei and energies of the
measured angular distributions are summarized in Table \ref{tab:iso}. 

The energies of the $^4$He$^{++}$ beam --- provided by the K20 cyclotron of ATOMKI --- were between 13.8 and 20.0 MeV with typical beam
currents of 300-500 enA. Since the imaginary part of the optical potential
depends sensitively on the energy, it is important to have a well-defined beam energy. Therefore the beam
was collimated by tight slits (1 mm wide) at the analyzing magnet; this  
corresponds to an overall energy spread of around 100 keV.

The targets were produced by evaporating isotopically highly enriched material ($\geq 95\%$, see Table \ref{tab:iso})
onto thin carbon foils ($\approx$ 20 $\mu$g/cm$^{2}$). The target and backing thicknesses were determined via alpha particle energy 
loss measurement using a mixed alpha source which contains $^{239}$Pu, $^{241}$Am, and $^{244}$Cm radioactive nuclides. Typical energy losses in the carbon backing and the targets are about 15 keV and within 40 and 70 keV, respectively. These energy 
losses correspond to about 150-200 $\mu g$/cm$^2$ target thicknesses. The targets were
mounted on a remotely controlled target ladder in the center of the scattering chamber. The stability of the targets was monitored
continuously during the experiment using detectors (see below) built into the wall of the scattering chamber at $\pm$ 15$^{\circ}$ (with respect to the beam direction).

\subsection{Scattering chamber, detectors, and angular calibration}

For carrying out the alpha elastic scattering experiments the 78.8 cm diameter scattering chamber was used.
Ion implanted silicon detectors with active areas of 50 mm$^2$ and 500 $\mu$m thickness have been used
to measure the yield of the scattered alpha particles. Two detectors were
mounted on the wall of the scattering chamber at fixed angles $\vartheta = \pm$15$^\circ$ left and right to the beam axis. These
detectors were used as monitor detectors --- their solid angles were 1.1 x 10$^{-6}$  --- during the whole experiment to normalize
the measured angular distribution and to determine the precise position of the beam on the target.
The detectors used to measure the alpha scattering angular distributions were mounted onto two independently movable turntables. On each of the turntables 
between 2 and 5 detectors were placed.
The solid angles of these detectors varied between 1.36 x 10$^{-4}$  and 1.93 x 10$^{-4}$ (calculated from the known geometry), and 
the ratio of these solid angles was determined precisely by measurements at overlapping angles with good
statistics ($\leq$1\% uncertainty). 

In order to derive precisely the scattering angle, the exact position of the beam on the target has to be known. 
For this reason, the following procedure was used. An aperture of 2 mm width and 6 mm height was mounted on the target
ladder to check the beam position and size of the beam spot before and after
every change of the beam energy or current. The beam was optimized until not
more than 1\% of the total beam current could be measured on this aperture. As
a result of the procedure, the horizontal size of the beam spot was below 2 mm
during the whole experiment.

The elastic scattering cross section at forward angles differs by several orders of magnitude from the one
measured at backward angles (see below), this fact necessitates very different counting times at forward and backward angles. 
While at forward angles the spectra were collected for typically 5-30 minutes,
at backward angles to achieve reasonable statistics the measuring times had to be typically within 3 and 6 hours. 
Furthermore, a reliable dead-time correction is also crucial. For this reason
the automatically determined dead-time provided by 
the data acquisition system has been verified using a pulser in all spectra.

Knowledge of the exact angular position of the detectors is of crucial
importance for the precision of a scattering experiment since the Rutherford
cross section depends sensitively on the scattering angle especially at
forward directions. The 
uncertainty in the angular distribution is dominated by the error of the
scattering angles in the forward region. 
To determine the scattering angle precisely, 
kinematic coincidences were measured between elastically scattered
alpha particles and the corresponding $^{12}$C recoil nuclei using a pure
carbon backing as target. One detector was placed at a certain angle to measure 
the yield of the alpha particles scattered on $^{12}$C, and its signal was
selected as a gate for the other detector which moved around the expected
$^{12}$C recoil angle. 
A Gaussian had been fitted onto this experimental yield data.
From the difference between the expected
 $^{12}$C recoil angle and the maximum of the Gaussian the angular
 offsets of the detectors can be calculated.
In addition, in some experiments
the steep kinematics of $^1$H\raa $^1$H scattering was analyzed
using a thin plastics (CH$_2$)$_n$ target.

This process was repeated for all detector pairs.
The angular offset  of each detector was less than  0.17$^\circ$ and 
the final angular uncertainty was found to be $\Delta$$\vartheta$ $\leq$
0.12$^\circ$ in the case of each detector. The given uncertainties of the
measured cross sections contain the uncertainty of the scattering angle using
standard error propagation; thus, the uncertainties of the scattering angle
are not given explicitly in the tables.

\subsection{Data analysis}

The elastic scattering peaks have to be well separated from the inelastic events and from alpha scattering on carbon (backing), oxygen, and other impurities of the target and/or backing. The energies of the first excited states of the target nuclei 
are listed in Table \ref{tab:iso}. In principle, if the separation between the elastic and inelastic scattering is not properly done, it can lead to an overestimation of the elastic scattering peak. However, since our energy resolution is about 60-80 keV and the energies of the first excited states are far above 500 keV, typically around 1 MeV, the separation could be properly done. 
Furthermore, the peaks corresponding to elastic and/or inelastic alpha
scattering on the carbon backing or on its oxygen contamination are also well
separated from the elastic alpha scattering events of our interest because of
the strong kinematic mass dependence of the energies of the scattered particles.

Angular distributions between 20$^{\circ}$ and 170$^{\circ} - 175^{\circ}$
were measured at each energy listed in Table \ref{tab:iso} (except for
$^{92}$Mo at 15.69 MeV; due to technical problems at the end of the
experiment, in this case the angular distribution was measured between
19.7$^{\circ}$ and 154.7$^{\circ}$, and the very backward angles could not be
covered), 
at angles below 100$^{\circ}$ in 1$^\circ$ steps and at angles above
100$^{\circ}$ in $2^\circ - 2.5^\circ$ angular steps. 
The statistical uncertainties varied typically between below 1\% (forward
angles) and 4\% 
(backward angles). The count rates \textit{N($\vartheta$)} have been
normalized to the yield of the monitor detectors
\textit{N$_{Mon.}$($\vartheta$=15$^\circ$)}: 
\begin{equation}
\left(\frac{d\sigma}{d\Omega}\right)(\vartheta)\,=\left(\frac{d\sigma}{d\Omega}\right)_{Mon.}\frac{N(\vartheta)}{N_{Mon.}}\frac{\Delta\Omega_{Mon.}}{\Delta\Omega},
\end{equation}
with $\Delta$$\Omega$ being the solid angles of the
detectors, and $(\frac{d\sigma}{d\Omega})_{Mon.}$ is approximately given by
the Rutherford cross section. Finally, the
measured cross sections are converted to the center-of-mass system.

Whereas the Rutherford normalized cross section data cover about two orders of magnitude
between the highest (forward angles at energies below the Coulomb barrier) and the lowest
measured cross sections (backward angles at energies above the Coulomb barrier), the underlying
cross sections cover more than four orders of magnitude. Over this huge range almost the same 
accuracy of about 2-5\% total uncertainty could be achieved for each of the studied reactions. 
This error is mainly caused by the uncertainty of the
determination of the scattering angle in the forward region and from the
statistical uncertainty in the backward region. 

The absolute normalization is done in two steps. In a first step the absolute
normalization is taken from experiment, i.e.\ from the integrated beam
current, the solid angle of the detectors, and the thickness of the
target. This procedure has a relatively large uncertainty of the order of
$10$\,\% which is mainly based on the uncertainties of the target
thickness. In a second step a ``fine-tuning'' of the absolute normalization is
obtained by comparison to theoretical calculations at very forward angles. It
is obvious that calculated cross sections from any reasonable potential
practically do not deviate from the Rutherford cross section at the most forward
angles; typical deviations are below 0.5\,\% for all
potentials. This ``fine-tuning'' changed the first experimental
normalization by only few per cent and thus confirmed the first normalization
within the given errors.

\section{Optical model analysis}
\label{sec:opt}
The comprehensive analysis of the elastic scattering angular distributions is carried out
within the framework of the optical model (OM). The interaction between the
\al\ projectile and the target nucleus is described by a complex OM potential
\begin{equation}
U_{\rm{OM}}(r) = V(r) + iW(r) + V_C(r)
\label{eq:pot}
\end{equation}
where $V(r)$, $W(r)$, and $V_C(r)$ are the real part, imaginary part, and
Coulomb part of the potential $U_{\rm{OM}}(r)$. The real part is calculated
from the double-folding model with the widely used DDM3Y interaction
\cite{Sat79,Kob84,Abe93,Atz96,moh97} where the folding potential $V_F(r)$ is
modified by two parameters:
\begin{equation}
V(r) = \lambda \, V_F(r/w) \quad \quad .
\label{eq:pot_real}
\end{equation}
$\lambda$ is the strength parameter of the order of $\lambda \approx 1.1 -
1.4$ leading to real volume integrals\footnote{As
  usual, the negative signs of the volume integrals $J_R$ and $J_I$ are
  omitted in the discussion.} of about $J_R \approx 350$\,MeV\,fm$^3$. $w
\approx 1$ is the width parameter which 
should remain very close to unity within about $1-2$\,\%; a larger deviation
of $w$ from unity would indicate a failure of the underlying folding
model. The folding potential is given by
\begin{equation}
V_F(r) = 
	\int \int \, \rho_P(r_P) \,\rho_T(r_T) \,
        v_{\rm{eff}}(s,\rho,E_{\rm{NN}}) \; d^3r_P \; d^3r_T
\label{eq:fold}
\end{equation}
with the nucleon densities $\rho_{P,T}(r_{P,T})$ and an effective
energy- and density-dependent nucleon-nucleon (NN) interaction $v_{\rm{eff}}$
which factorizes into 
\cite{Kob84}
\begin{equation}
v_{\rm{eff}}(|\vec{s}|,\rho_P,\rho_T,E_{\rm{NN}}) = 
	f(|\vec{s}|,E_{\rm{NN}}) \cdot g(\rho_P,\rho_T,E_{\rm{NN}}) \quad
        \quad .
\label{eq:v_eff}
\end{equation}
The energy $E_{\rm{NN}}$ in the NN interaction is usually taken from the
energy per nucleon of the \al\ projectile: $E_{\rm{NN}} = E_{\rm{lab}}/A_P$.
$f(|\vec{s}|,E_{\rm{NN}})$ is composed of a short-range repulsive and a
longer-range attractive part and a zero-range exchange part
(e.g.\ \cite{Sat79}): 
\begin{equation}
f(|\vec{s}|,E_{\rm{NN}}) = \Bigl[
	7999 \, \frac{e^{-4s}}{4s} - 2134 \, \frac{e^{-2.5s}}{2.5s}
	- 276 \, ( 1 - 0.005\,E_{\rm{NN}} ) \, \delta(s) \Bigr]~{\rm{MeV}}
\label{eq:nn_f}
\end{equation}
The geometrical definition of the interaction distance $s$ in
Eq.~(\ref{eq:nn_f}) is shown e.g.\ in Fig.~1 of \cite{Abe93} ($s$ in fm). The
density-dependent $g(\rho_P,\rho_T,E_{\rm{NN}})$ is given by
\begin{equation}
g(\rho_P,\rho_T,E_{\rm{NN}}) = C(E_{\rm{NN}}) \cdot 
	\left[ 1 + \alpha(E_{\rm{NN}}) \cdot 
          e^{-\beta(E_{\rm{NN}})\cdot(\rho_P+\rho_T)} \right]
\label{eq:nn_g}
\end{equation}
within the so-called frozen-density approximation.
The parameters $C(E_{\rm{NN}})$, $\alpha(E_{\rm{NN}})$, and
$\beta(E_{\rm{NN}})$ of the density dependence 
in Eq.~(\ref{eq:nn_g}) are fitted to reproduce the strength of an effective
G-matrix interaction \cite{Jeu77,Jeu81}. The results for the parameters $C$,
$\alpha$, and $\beta$ as adopted in \cite{Abe93} are listed in Table
\ref{tab:interaction}. 

It has to be noted that the influence of the chosen parameters for the
density dependence is minor for several reasons. First of all, the energy
dependence of the parameters $C$, $\alpha$, and $\beta$ is minor: in the shown
energy range in Tab.~\ref{tab:interaction} (5\,MeV $\le E_{\rm{lab}} \le$
40\,MeV) the parameters $C$, $\alpha$, and $\beta$ do not vary by more than
about 5\,\%; the energy range of the scattering experiments in this study is
much smaller. 
Second, the parameter $C$ scales the total strength of the
interaction; any arbitrary choice of the parameter $C$ will be compensated
during the fitting procedure by the adjustment of the potential strength
parameter $\lambda$. The parameters $\alpha$ and $\beta$ have some influence
on the shape of the folding potential; however, this is at least partly
compensated by fitting the potential width parameter $w$. Instead of
discussing the parameters $\lambda$ and $w$ it is recommended to discuss the
real volume integrals $J_R$ and root-mean-square radii $r_{R,rms}$ of
the fitted real potential because these parameters are practically independent
of the chosen parameters $C$, $\alpha$, and $\beta$ of the density
dependence. 

The basic ingredients in the calculation of the real folding potential are the
nucleon densities of the colliding nuclei. They are derived from measured
charge density distributions which are compiled in \cite{Vri87}. For the \al
-particle the sum-of-Gaussian parameterization was chosen because the
corresponding measurement covers by far the largest range of momentum
transfer. For the target nuclei the average potential from all available
density parameterizations was used. For $^{106}$Cd the two-parameter Fermi
distribution of $^{110}$Cd (with $R_0$ scaled by $A^{1/3}$) had to be used
because no data for $^{106}$Cd are listed in \cite{Vri87}. The variations of
$R_0$ along the cadmium isotopic chain are of the order of 0.5\,\%; thus, the
uncertainty of the extrapolated density distribution should remain small.

$W(r)$ is taken as Woods-Saxon potential (surface type):
\begin{equation}
W(r) = W_S \times \frac{df_{WS}(x_S)}{dx_S}
\label{eq:imag}
\end{equation}
with the potential depth\footnote{Note that maximum depth of the imaginary
  potential $W(r)$ at $r = R_S \times
  A_T^{1/3}$ is $W_{\rm{max}}(r) = -W_S/4$ in the
  definition of Eq.~(\ref{eq:imag}).} 
$W_S$ and 
\begin{equation}
f_{WS}(x_S) = \frac{1}{1+\exp{(x_S)}}
\end{equation}
and $x_S = (r-R_S \times A_T^{1/3})/a_S$ with the radius parameter $R_S$ and the
diffuseness parameter $a_S$. It has been found in many studies that the
surface contribution is dominating at low energies, and the volume part is
small or even 
negligible \cite{gal05,Mohr11}; this finding has recently been confirmed in a
microscopic calculation of the OM potential \cite{Guo11}. Finally, the Coulomb
potential $V_C(r)$ has been calculated from a homogeneously charged sphere with
a Coulomb radius $R_C$ identical to the root-mean-square radius $r_{R,rms}$ of
the real folding potential (before adjusting the width parameter,
i.e.\ $w=1$). 

The above choice of the potential reduces the number of adjustable
parameters. Three parameters of the imaginary part ($W_S$, $R_S$, $a_S$) and
two parameters of the real part ($\lambda$, $w$) were fitted to the angular
distributions using a standard $\chi^2$ search. However, because of
ambiguities of the potential the parameter search was restricted to the above
mentioned region of real volume integrals $J_R \approx 350$\,MeV\,fm$^3$. 
The ambiguities increase
towards lower energies, and in some cases very shallow minima in $\chi^2$ have
been found with parameters $w$ which deviated by more than 3\,\% from unity. In
such cases, the width parameter $w$ has been fixed either at $w = 1.0$ or at
the average value found at the higher energies (note that typically this
average value does not deviate by more than 1\,\% from unity). The fits are
shown in Graphs~\ref{fig:y89mo92}, \ref{fig:cd106cd110}, and
\ref{fig:cd116sm144}, and the parameters are listed in Table 
\ref{tab:pot}.

In addition, the total reaction cross section \stot\ has been calculated from
the obtained potential by the well-known formula
\begin{equation}
\sigma_{\rm{reac}} =
   \frac{\pi}{k^2} \sum_L (2L+1) \, (1 - \eta_L^2)
\label{eq:stot}
\end{equation}
where $k = \sqrt{2 \mu E}/\hbar$ is the wave number,
$E$ is the energy in the center-of-mass (c.m.)\ system, and
$\eta_L$ are the real reflexion coefficients. The uncertainties of \stot\ have
been estimated from phase shift fits using the model of \cite{Chi96} and are
typically of the order of about 3\,\% (except at the lowest energies); see
also the discussion of uncertainties of \stot\ in \cite{Mohr10}. Minor
differences 
between \stot\ in this work and our previous study \cite{Mohr10} -- typically
within the uncertainties of this work -- are due to the fact that \stot\ in
\cite{Mohr10} are calculated from different potential parameterizations (volume
and surface Woods-Saxon in the imaginary part) whereas this work uses an
imaginary Woods-Saxon potential of pure surface type.

\begin{figure}[ht!]
\centering
\includegraphics{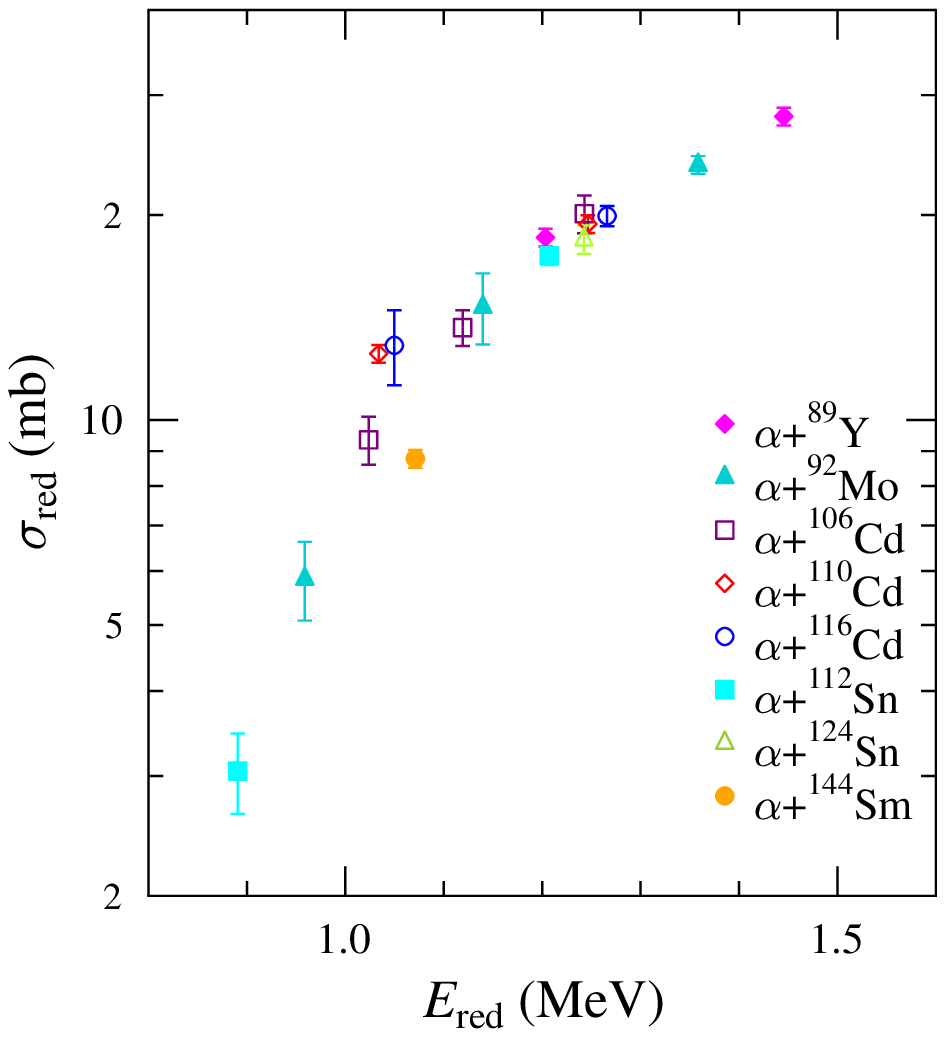}
\caption{Reduced cross section \sred\ versus the reduced energy \ered\
  for the data in Table \ref{tab:pot} and Graphs~\ref{fig:y89mo92},
  \ref{fig:cd106cd110}, and \ref{fig:cd116sm144}.}
\label{fig:sigred}
\end{figure}

For the comparison of total reaction cross sections \stot\ for various
projectile-target systems at different energies it has been suggested to
present the data as reduced cross sections \sred\ versus the reduced energy
\ered\ which are defined by
\begin{eqnarray}
E_{\rm{red}} & = & \frac{\bigl(A_P^{1/3}+A_T^{1/3}\bigr) E_{\rm{c.m.}}}{Z_P Z_T} \\
\sigma_{\rm{red}} & = & \frac{\sigma_{\rm{reac}}}{\bigl(A_P^{1/3}+A_T^{1/3}\bigr)^2}
\label{eq:red}
\end{eqnarray}
The reduced energy \ered\ takes into account the different heights of
the Coulomb barrier in the systems under consideration, whereas the reduced
reaction cross section \sred\ scales the measured total reaction cross section
\stot\ according to the geometrical size of the projectile-plus-target
system. It is found that the reduced cross sections \sred\ show a very similar
behavior for all nuclei under study (see Fig.~\ref{fig:sigred}). Significantly
different \sred\ are found for weakly bound projectiles (like
e.g.\ $^{6,7}$Li) or halo projectiles (e.g.\ $^6$He), see \cite{Far10}.

The results of this section are used as the basis for the construction of a
new systematic \al -nucleus optical potential with very few adjustable
parameters. Further details of the new potential are given in the Appendix
(Sect.\ref{sec:new}).

\section{Summary and Conclusions}
\label{sec:summ}
The extraction of optical potentials from elastic scattering angular
distributions requires experimental data with high precision. This holds in
particular at relatively low energies around the Coulomb barrier where the
repulsive Coulomb interaction governs the measured angular
distribution. Increasing ambiguities of the derived potentials are found when
the energy is lowered below the Coulomb barrier, e.g.\ when approaching
astrophysically relevant energies.

Here we have presented a summary of the experimental scattering data for the
nuclei $^{89}$Y, $^{92}$Mo, $^{106,110,116}$Cd, $^{112,124}$Sn, and $^{144}$Sm
which were measured at ATOMKI, Debrecen in the last 15 years. The target nuclei
were chosen to cover even-even and even-odd nuclei with magic and non-magic
proton and neutron numbers. The measured angular distributions cover the full
angular 
range from about $20^\circ$ to $170^\circ$ in small steps of $1^\circ -
2^\circ$ with uncertainties of a few per cent. These data are used to extract
locally adjusted optical potentials, and from the obtained local potential
parameters a new \al -nucleus potential is suggested which is able to
reproduce the experimental total reaction cross section \stot\ at the measured
energies with deviations of less than 5\,\% in most cases and less than 20\,\%
in the worst case.

Elastic scattering data are the indispensable basis for the extraction of the
optical potential, but this basis is unfortunately not complete yet. Angular
distributions are available for targets in the mass 
range around $A \approx 100$, but complete angular distributions with small
uncertainties are still missing for heavier targets above $A \approx 150$.
There are significant complications of the analysis of scattering
data at energies below the Coulomb barrier. Thus, in addition to the analysis
of elastic scattering angular distributions, further experimental cross
section data for \al -induced reactions are required to
determine the weak energy dependence of the real part and the strong energy
dependence of the imaginary part of the optical potential. Such a combined
effort using elastic scattering and reaction cross sections
should be able to resolve or at least to reduce the long-standing
problem of \al -nucleus potentials at low energies and the resulting
uncertainties in the prediction of \al -induced reaction cross sections in the
near future.

\ack

This work was supported by the EUROGENESIS research program, by OTKA (NN83261,
K101328), by the European Research Council (grant agreement no. 203175). The
authors thank for the collaboration to M.\ Avrigeanu, M.\ Babilon, Z.\ Elekes,
J.\ Farkas, J.\ G\"orres, R.\ T.\ G\"uray, M.\ Jaeger, Z.\ M\'at\'e,
A.\ Kretschmer, S.\ M\"uller, H.\ Oberhummer, A.\ Ornelas,
N.\ \"Ozkan, T.\ Rauscher, K.\ Sonnabend,
G.\ Staudt, C.\ Yalcin, A.\ Zilges, and L.\ Zolnai, and to the ATOMKI
cyclotron staff for the stable beam in all the experiments.

\appendix
\section{Appendix: Suggestions for a new \al -nucleus potential}
\label{sec:new}
In the following we present some suggestions for a new systematic \al -nucleus
optical potential which is derived from the ATOMKI scattering data in this
study. At present the potential is based on elastic scattering data in the
mass range $89 \le A \le 144$ and in the energy range 13\,MeV $\lesssim E
\lesssim $ 20\,MeV. The mass range can be extended in a standard way by
measuring and analyzing angular distributions for targets with $A < 89$ and $A
> 144$. As has been pointed out in \cite{Atz96}, the \al -nucleus potential is
not very sensitive to the mass of the target above $A \approx 60$;
consequently, the new potential may be considered as valid for $A > 60$. An
extension of the energy range to higher energies is not planned because it has
been shown (see e.g.\ \cite{Mohr11,gal05,Guo11}) that the imaginary part of the
potential changes from a dominating surface potential at low energies to a
dominating volume potential at energies above $\approx 30 - 40$\,MeV. An
extension to lower energies is essential for the prediction of low-energy
cross sections of \al -induced reactions. Such an extension cannot be done 
easily by
elastic scattering because the cross section approaches the Rutherford cross
section. Instead, experimental reaction cross sections have to be used to
further constrain the \al -nucleus potential. It is interesting to note that
under certain conditions the reaction cross section of a particular
\raX\ reaction is almost entirely defined by the \al -nucleus potential.

The following suggestions for the new systematic potential are based on two
new ideas. 
First, as a careful inspection of Table \ref{tab:pot} shows, there is only a
relatively weak variation of the potential parameters for all nuclei and all
energies under study. Obviously such a finding is very helpful in the
construction of a global potential. However, the relatively small variation of
the potential parameters turns out to be non-systematic (except the increase
of the imaginary volume integral $J_I$ with increasing energy) which
complicates the determination of a global potential. We will
accommodate this finding by a simple averaging procedure for the potential
parameters as outlined below to keep the number of adjustable parameters as
low as possible. Of course, a many-parameter global potential (as
e.g.\ suggested in \cite{Avr10,Avr09,Pal08,Pal12}) will be able to reproduce
elastic angular distributions with smaller deviations in $\chi^2$; however,
because of the non-systematic behavior of the potential parameters any
extrapolation using many-parameter global potentials may become uncertain.

\begin{figure}[ht!]
\centering
\includegraphics{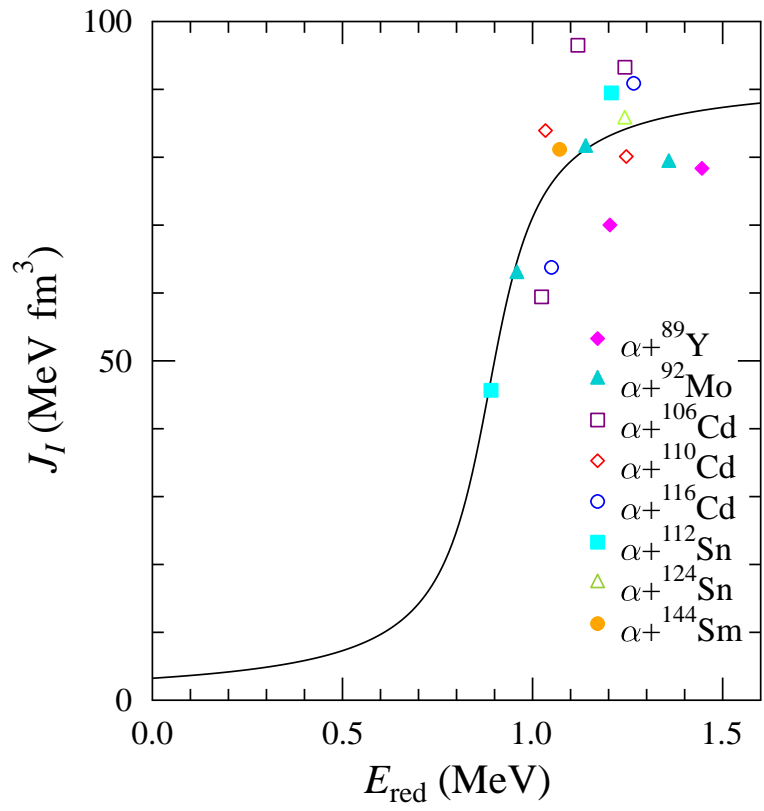}
\caption{Imaginary volume integrals $J_I$ versus the reduced energy \ered
  . Further discussion see text.}
\label{fig:ji}
\end{figure}

Second, we suggest a new energy dependence of the imaginary volume
integral. This suggestion is based on a new parametrization of the energy
dependence of $J_I$ as a function of the reduced energy \ered\ (see
Fig.~\ref{fig:ji}).

We place special emphasis on the reproduction of the total reaction cross
section \stot\ because \stot\ is the essential basis of any calculation of
reaction cross sections
within the statistical model. As will be shown, the deviation
between the experimental \stot\ and the prediction from the suggested
potential remains small in all cases under study.

Of course, the derived parameters 
and its uncertainties may be further improved by including more scattering
data as e.g.\ from \cite{Mohr11,Pal08,Pal12}. 
The suggested potential has to be tested
against experimental cross sections of \al -induced reactions, in particular
\rag , \ran , and \rap\ reactions. Such tests and further optimizations of
this global potential are beyond the scope of the
present paper and will be presented elsewhere.

The geometry of the real part of the potential can be determined from the
folding procedure with very small uncertainties. From all angular
distributions we find an average width parameter $w_{\rm{all}} = 1.0061 \pm
0.0095$. The analysis of all semi-magic (SM) target nuclei results in
$w_{\rm{SM}} = 1.0104 \pm 0.0108$, and from the non-magic (NM) target nuclei
we find $w_{\rm{NM}} = 1.0006 \pm 0.0025$. Thus, it is reasonable to adopt $w
= 1.0$ for the global potential. It has to be noted that there are
still significant uncertainties for the real part of the \al -nucleus
potentials for unstable 
nuclei because density distributions from electron scattering are only
available for stable nuclei \cite{Vri87}.

The energy dependence of the real part of the \al -nucleus potential is
expected to be very weak at low energies. The intrinsic energy dependence of
the interaction, see Eqs.~(\ref{eq:nn_f}) and (\ref{eq:nn_g}) above, leads to
decreasing volume integrals $J_R$ with increasing energy. However, the
coupling of the real and imaginary part by a dispersion relation leads to an
opposite effect at low energies. No systematic trend for the energy
dependence of the volume 
integrals $J_R$ can be found in our scattering data. Thus, we adopt an
energy-independent strength of the real part of our global potential.

The strength parameter $\lambda$ of the real part has to be derived from the
energy-independent volume integral $J_R$. Here we find the average values
of 
$J_{R,{\rm{all}}} = 354.9 \pm 20.8$\,MeV\,fm$^3$, 
$J_{R,{\rm{SM}}} = 342.4 \pm 9.6$\,MeV\,fm$^3$, and
$J_{R,{\rm{NM}}} = 371.0 \pm 20.6$\,MeV\,fm$^3$.
Because of the differences between $J_{R,{\rm{SM}}}$ and $J_{R,{\rm{NM}}}$ we
recommend to use the lower value of 342.4\,MeV\,fm$^3$ for semi-magic targets
and the higher value of 371.0\,MeV\,fm$^3$ for non-magic targets. 

It is more difficult to provide a reasonable imaginary part for a global \al
-nucleus potential because it is well-known that the imaginary part increases 
with energy because of the increasing number of open channels. 
(Note that the total reaction cross section \stot\ increases strongly with
energy even if there is no energy dependence of the imaginary potential; this
is simply due to the increasing tunneling probablility at higher energies.)
We
find from our scattering data that the geometry of the imaginary part is well
constrained by the scattering angular distributions. The average radius
parameters $R_S$ of the imaginary surface Woods-Saxon potential are 
$R_{S,{\rm{all}}} = 1.430 \pm 0.086$\,fm,
$R_{S,{\rm{SM}}} = 1.477 \pm 0.038$\,fm, and
$R_{S,{\rm{NM}}} = 1.370 \pm 0.094$\,fm; the average diffuseness values are
$a_{S,{\rm{all}}} = 0.470 \pm 0.100$\,fm,
$a_{S,{\rm{SM}}} = 0.423 \pm 0.056$\,fm, and
$a_{S,{\rm{NM}}} = 0.531 \pm 0.114$\,fm. Because of the relatively small
deviations between the results for semi-magic and non-magic nuclei we adopt
the average values from all scattering data: $R_S = 1.430$\,fm and $a_S =
0.470$\,fm. 

The strength of the imaginary part is parametrized by the imaginary volume
integral $J_I$. Here we find a relatively well-defined behavior for the $J_I$
vs.\ \ered\ dependence which can be fitted well by the following formula
(similar to the phase shift behavior in a resonance):
\begin{equation}
J_I(E_{\rm{red}}) = \frac{1}{\pi} \, J_{I,0} \,
\arctan{\frac{\Gamma_{\rm{red}}}{2(E_{{\rm{red}},0} - E_{\rm{red}})}}
\label{eq:ji}
\end{equation}
with the saturation value $J_{I,0} = 92.0$\,MeV\,fm$^3$, the turning point
energy $E_{{\rm{red}},0} = 0.89594$\,MeV, and the slope (or width) parameter
$\Gamma_{\rm{red}} = 0.19659$\,MeV. The data and the fit are shown in
Fig.~\ref{fig:ji}. The parametrization in Eq.~(\ref{eq:ji}) has been chosen
because the widely used Brown-Rho parametrization of $J_I$ \cite{Bro81} has
turned out to be inadequate for the description of low-energy reaction cross
sections (see e.g.\ \cite{Som98,Mohr11}). A Fermi-type function for $J_I$
(with the parameters as e.g.\ used in \cite{Som98,Sau11,Mohr11}) has the
disadvantage that $J_I$ almost vanishes at energies below 10\,MeV, and thus
the calculated total cross section \stot\ becomes smaller than the inelastic
Coulomb excitation cross section. Eq.~(\ref{eq:ji}) is some kind of
compromise between the intrinsically steep $J_I(E)$ from a Fermi-type function 
and the intrinsically shallow $J_I(E)$ from a Brown-Rho parametrization.

In the considered mass range $89 \le A \le
144$ the depth $W_S$ of the imaginary surface Woods-Saxon potential for the
chosen geometry ($R_S = 1.43$\,fm, $a_S = 0.47$\,fm)
is approximately related to the volume integral $J_I$ by
\begin{equation}
W_S \approx (0.8112 + 0.008363 \, A_T - 1.432 \times 10^{-5}\, A_T^2) \times J_I
\label{eq:ws}
\end{equation}
with $W_S$ in MeV and $J_I$ in MeV\,fm$^3$ in Eq.~(\ref{eq:ws}); this equation
holds also for $50 \le A \le 210$ with small numerical deviations. The results
from this new systematic potential are shown in Graphs~\ref{fig:y89mo92},
\ref{fig:cd106cd110}, and \ref{fig:cd116sm144} as dashed lines, and the
resulting deviations $\chi^2/F$ and total reaction cross sections \stot\ are
listed in Table \ref{tab:glob}. It is obvious that the global potential
deviates stronger from the experimental data than the locally fitted
potential. Nevertheless, the experimental total reaction cross sections
\stot\ are well reproduced in all cases.

Here we provide a new \al -nucleus potential with a very limited
number of parameters. The folding potential in the real part has to be scaled
to the corresponding volume integral $J_R$ for semi-magic or non-magic
targets (i.e., one parameter for the real part). The imaginary surface
Woods-Saxon potential has an energy-independent geometry (i.e., two parameters
for the imaginary geometry $R_S$ and $a_S$),
and its energy-dependent strength is parametrized by the saturation value
$J_{I,0}$, the turning point energy $E_{{\rm{red}},0}$, and the slope 
parameter $\Gamma_{\rm{red}}$ (i.e., three parameters for the imaginary energy
dependence). The parameters for the calculation of the depth $W_S$ in
Eq.~(\ref{eq:ws}) do not count here, because these parameters are only a
technical help to calculate $W_S$ from a given $J_I$.
This small number of parameters is close to the widely used 4-parameter
potential by McFadden and Satchler \cite{McF66}, but significantly smaller
than the recent many-parameter potentials by Avrigeanu {\it et
  al.}~\cite{Avr10} or Palumbo {\it et al.}~\cite{Pal12}. It follows the idea
of \cite{Avr10} that the energy dependence of the imaginary part should be
related to the height of the Coulomb barrier because the imaginary volume
integrals $J_I$ are parametrized in dependence of the reduced energy \ered .

The suggested new potential shows relatively poor $\chi^2$ for $^{124}$Sn
and $^{144}$Sm and is also not perfect for $^{110,116}$Cd. The problems with
$^{124}$Sn and $^{144}$Sm may simply be related to the fact that the derived
average parameters are not very sensitive to the measurements of $^{124}$Sn
and $^{144}$Sm because in these cases only one energy has been measured. It is
obvious that the average parameters are most influenced by the many data
points from the lower-mass target nuclei with $A \approx 100$. This clearly
calls for further experimental data for heavier targets with masses far above
$A \approx 100$, i.e.\ in the mass range $150 \le A \le 210$.

There is no simple explanation why the new global potential works well for
$^{106}$Cd, but shows some problems with $^{110}$Cd and $^{116}$Cd. The
angular distributions for $^{110}$Cd and $^{116}$Cd look similar (with about
$\sigma/\sigma_R \approx 0.015$ at very backward angles in the 19\,MeV angular
distribution) whereas the elastic
scattering cross section for $^{106}$Cd is almost a factor of two larger
($\sigma/\sigma_R \approx 0.03$). It has to be expected that any global
potential with few smoothly varying parameters will show similar problems with
the description of angular distributions along the cadmium isotopic chain.

Although this simple global potential is not perfect in the reproduction of
the angular distributions, it nevertheless reproduces the total reaction cross
sections \stot\ within the given uncertainties in most cases. Even in the
worst case ($^{110}$Cd, $E = 15.56$\,MeV) the deviation does not exceed 20\,\%
for \stot ; such an agreement is the basic prerequisite for the prediction of
\al -induced reaction cross sections. However, further studies beyond this
paper are clearly necessary; the presented new potential should be
considered as a first version which is based on the scattering data 
of this work only.  The
largest uncertainty of this global potential results from the parametrization
of the energy dependence
of the imaginary volume integral $J_I(E)$ in Eq.~(\ref{eq:ji}). Several
parametrizations have been suggested in the literature for the energy dependence
of $J_I$ but a strict theoretical derivation of this energy dependence is
missing. In addition, the parameters of any $J_I$-vs.-\ered\ dependence cannot
be fully constrained by elastic scattering experiments which are typically
done only at energies around and above the turning point of Eq.~(\ref{eq:ji}),
but not below; see also
Fig.~\ref{fig:ji}. The saturation value $J_{I,0} = 92$\,MeV\,fm$^3$ is well
defined from the elastic scattering data around the Coulomb barrier with about
$10-15$\,\% uncertainty; this uncertainty has only minor influence on the
prediction of reaction cross sections below the Coulomb barrier. The
uncertainty of the turning point energy $E_{{\rm{red}},0}$ is also of the
order of 10\,\%; but together with the much larger uncertainty of the slope
parameter $\Gamma_{\rm{red}}$ this leads to considerable uncertainties for
$J_I(E)$ at energies below the Coulomb barrier which may exceed a factor of
two or three at reduced energies below or around $E_{\rm{red}} \approx 0.7$\,MeV
which corresponds to energies of e.g.\ about 9\,MeV for $^{89}$Y and about
13\,MeV for $^{144}$Sm. This uncertainty of $J_I(E)$ has significant impact on
the prediction of reaction cross sections at low energies. Consequently,
reaction data have to be taken into account for better
constraining these parameters $E_{{\rm{red}},0}$ and $\Gamma_{\rm{red}}$
in particular at lower energies. Furthermore, as
pointed out above, the radius $R_S$ and diffuseness $a_S$ seem to be slightly
different for semi-magic and non-magic target nuclei. Of course it is also
possible that the parameters $J_{I,0}$, $\Gamma_{\rm{red}}$, and
$E_{{\rm{red}},0}$ in Eq.~(\ref{eq:ji}) depend on whether the proton or
neutron number of the target nucleus is magic or not. Such investigations will
be presented in a forthcoming study.



\clearpage

\section*{Figures}

\clearpage




\bigskip


\bigskip

\newpage

\renewcommand{\baselinestretch}{1.0}

\begin{center}
\begin{table*}
\caption{\label{tab:iso} 
  Charge ($Z$) and neutron ($N$) number, chemical form, and isotopic enrichment
  of the target material, $E_{\rm{c.m.}}$ 
  energies for each of the angular distributions studied at ATOMKI in the recent
  years.
Furthermore, the energies of the first excited states of the target nuclei 
are listed, too.} 
\begin{tabular}{ccccccccc}
\hline
\parbox[t]{0.6cm}{\centering{target \\ nuclei }} &
\parbox[t]{0.4cm}{\centering{$Z$}} &
\parbox[t]{0.4cm}{\centering{$N$}} &
\parbox[t]{1.2cm}{\centering{chemical \\ form}} &
\parbox[t]{1.2cm}{\centering{enrichment (\%) }} &
\parbox[t]{1.0cm}{\centering{$E_{\rm{c.m.}}$ (MeV) }} &
\parbox[t]{2.0cm}{\centering{1.st excited \\ state (keV)}} &
\parbox[t]{1.0cm}{\centering{Ref. }} \\
\hline
 $^{89}$Y &   39    &  50     & metallic & 100 &  15.51  & 908.97 & \cite{kis09} \\
                  &           &           &             &        &  18.63  &&                      \\
 $^{92}$Mo&  42   &   50    & oxide (MoO$_3)$&97.3 &  13.20  &1509.51& \cite{ful01} \\
                    &         &           &             &        &  15.69  &&                      \\
                    &         &           &             &        &  18.70  &&                      \\
 $^{106}$Cd & 48 & 58  & metallic & 96.5 & 15.55    &632.64 & \cite{gal_phd, kis06} \\
                      &      &       &              &         & 17.00  &&                     \\
                      &      &       &              &         & 18.88  &&                      \\
$^{110}$Cd & 48 & 62  & metallic & 95.7 &  15.56  &657.76 &  \cite{kis11}\\
                     &      &       &              &         &  18.76   &&                     \\
$^{116}$Cd & 48 & 68  & metallic & 98.3 &  15.59    &513.49 &  \cite{kis11} \\ 
                     &      &       &              &         &  18.80   && \\
 $^{112}$Sn & 50 & 62  & metallic & 99.6 & 13.90 &1256.85 &  \cite{gal05}\\ 
                      &      &       &              &         &  18.84   && \\
$^{124}$Sn  & 50 & 74  & metallic&  97.4&  18.90  &1131.74 &  \cite{gal05}\\ 
$^{144}$Sm & 62 & 82  & oxide (Sm$_2$O$_3)$ & 96.5&  19.45  &1660.03 &  \cite{moh97}\\
\hline
\end{tabular}
\end{table*}
\end{center}

\bigskip

\begin{table}
\begin{center}
\caption{
   \label{tab:pot} 
  Parameters of the optical potentials derived from the fits to the elastic
  scattering angular distributions (see Graphs~\ref{fig:y89mo92},
  \ref{fig:cd106cd110}, and \ref{fig:cd116sm144}) and the 
  total reaction cross section \stot\ (see also Fig.~\ref{fig:sigred}).
}
\begin{tabular}{ccccccrccccr@{$\pm$}lc}
\hline
target & $E$
& $\lambda$ & $w$ & $J_R$ & $r_{rms,R}$ 
& \multicolumn{1}{c}{$W_s$} & $R_S$ & $a_S$ & $J_I$ & $r_{rms,I}$ 
& \multicolumn{2}{c}{\stot } & $\chi^2/F$ \\
& (MeV) & ($-$) & ($-$) & (MeV\,fm$^3$) & (fm)
& \multicolumn{1}{c}{(MeV)} & (fm) & (fm) & (MeV\,fm$^3$) & (fm) 
& \multicolumn{2}{c}{(mb)} & ($-$) \\
\hline
$^{89}$Y & 15.51
& 1.286 & 1.001	& 341.0 & 4.967 
& 100.8	& 1.457	& 0.457	& 70.0 & 6.763 & 678.9 & 20.4 & 1.56 \\
& 18.63	
& 1.214	& 1.016	& 337.0	& 5.042	
& 107.7 & 1.481	& 0.464	& 78.4 & 6.871 & 1022.4	& 30.7 & 3.31 \\
$^{92}$Mo & 13.20
& 1.259 & 1.010 & 345.1 & 5.054
& 101.6 & 1.517 & 0.382 & 62.8 & 7.020 & 217.7 & 28.8 &	2.25 \\
& 15.69	
& 1.190 & 1.008 & 323.7	& 5.041
& 126.5 & 1.464 & 0.426 & 81.4 & 6.828 & 545.9 & 65.4 & 4.62 \\
& 18.70	
& 1.217 & 1.013 & 336.8 & 5.071 
& 119.2 & 1.486	& 0.427 & 79.2 & 6.925 & 882.3 & 26.5 &	3.49 \\
$^{106}$Cd & 15.55
& 1.314 & 1.002 & 350.7	& 5.220	
& 87.3	& 1.474	& 0.465	& 59.4 & 7.225 & 373.6 & 30.4 &	2.07 \\
& 17.00	
& 1.373 & 1.004	& 368.9	& 5.232
& 166.3	& 1.464	& 0.403 & 96.5 & 7.116 & 546.0 & 33.1 &	0.87 \\
& 18.88	
& 1.379 & 1.000 & 365.3 & 5.208
& 138.7	& 1.402	& 0.505	& 93.3 & 6.942 & 802.3 & 51.3 &	0.77 \\
$^{110}$Cd & 15.56
& 1.560 & 1.000 & 412.7 & 5.251	
& 120.5	& 1.205 & 0.699 & 83.9 & 6.423 & 509.1 & 15.3 & 0.70 \\
& 18.76	
& 1.387 & 0.996 & 362.9	& 5.232 
& 129.7	& 1.381	& 0.485 & 80.1 & 6.901 & 789.6 & 23.8 & 0.32 \\
$^{116}$Cd & 15.59
& 1.442 & 1.000	& 379.6	& 5.303 
& 82.8 & 1.298 & 0.684 & 63.8 & 6.905 &	537.8 & 68.0 & 0.27 \\
& 18.80	
& 1.347	& 1.002 & 356.6 & 5.313
& 157.3 & 1.366 & 0.472 & 90.9 & 6.931 & 833.1 & 28.5 & 0.61 \\
$^{112}$Sn & 13.90
& 1.326 & 1.000 & 352.9 & 5.293
& 61.8 & 1.536 & 0.474 & 45.7 &	7.646 &	125.3 & 17.0 & 0.61 \\
& 18.84	
& 1.354 & 0.996 & 356.5	& 5.273
& 146.9 & 1.423	& 0.455 & 89.5 & 7.101 & 715.1 & 21.5 &	0.63 \\
$^{124}$Sn & 18.90
& 1.251 & 1.020 & 344.9 & 5.444
& 153.0 & 1.428 & 0.430 & 85.6 & 7.330 & 795.8 & 38.5 &	1.12 \\
$^{144}$Sm & 19.45
& 1.207 & 1.030 & 343.7 & 5.742
& 204.8 & 1.499 & 0.293 & 81.2 & 7.944 & 409.2 & 12.3 &	1.82 \\
\hline
\end{tabular}
\end{center}
\end{table}

\bigskip

\begin{table}
\begin{center}
\caption{
   \label{tab:glob} 
  Results for $\chi^2$ and \stot\ from the simple global potential as
  suggested in Sec.~\ref{sec:new}. 
}
\begin{tabular}{ccr@{$\pm$}lrc}
\hline
& & \multicolumn{2}{c}{experiment} 
& \multicolumn{2}{c}{global potential} \\
target & $E$
& \multicolumn{2}{c}{\stot} 
& \multicolumn{1}{c}{$\chi^2/F$} & \stot \\
& (MeV) & \multicolumn{2}{c}{(mb)} 
& \multicolumn{1}{c}{($-$)} & \multicolumn{1}{c}{(mb)} \\
%
\hline
$^{89}$Y & 15.51
& 678.9 & 20.4 & 1.8 & 684.5 \\
& 18.63	
& 1022.4& 30.7 & 12.0 & 985.5 \\
$^{92}$Mo & 13.20
& 217.7 & 28.8 & 4.0 & 228.4 \\
& 15.69	
& 545.9 & 65.4 & 9.6 & 564.0 \\
& 18.70	
& 882.3 & 26.5 & 13.4 & 877.9 \\
$^{106}$Cd & 15.55
& 373.6 & 30.4 & 7.7 & 380.5 \\
& 17.00	
& 546.0 & 33.1 & 5.9 & 572.7 \\
& 18.88	
& 802.3 & 51.3 & 1.3 & 785.3 \\
$^{110}$Cd & 15.56
& 509.1 & 15.3 & 21.8 & 404.8 \\
& 18.76	
& 789.6 & 23.8 & 16.9 & 801.1 \\
$^{116}$Cd & 15.59
& 537.8 & 68.0 & 13.3 & 439.0 \\
& 18.80	
& 833.1 & 28.5 & 65.7 & 842.0 \\
$^{112}$Sn & 13.90
& 125.3 & 17.0 & 2.7 & 104.5 \\
& 18.84	
& 715.1 & 21.5 & 2.0 & 725.0 \\
$^{124}$Sn & 18.90
& 795.8 & 38.5 & 498.2 & 769.4 \\
$^{144}$Sm & 19.45
& 409.2 & 12.3 & 347.5 & 431.6 \\
\hline
\end{tabular}
\end{center}
\end{table}

\clearpage
\newpage

\TableExplanation

\bigskip
\renewcommand{\arraystretch}{1.0}

\section*{Table 1.\label{tbl1te} 
Parameters $C(E_{\rm{NN}})$, $\alpha(E_{\rm{NN}})$, and $\beta(E_{\rm{NN}})$
of the density dependence of the chosen nucleon-nucleon interaction}
\begin{tabular*}{0.95\textwidth}{@{}@{\extracolsep{\fill}}lp{5.5in}@{}}
$E_{\rm{lab}}$ & Energy of the $\alpha$ projectile in the laboratory system in
  MeV \\
$E_{\rm{NN}}$  & Energy per nucleon in MeV: $E_{\rm{NN}} = E_{\rm{lab}}/A_P$ \\
$C(E_{\rm{NN}})$ & Coefficient $C$ in Eq.~(\ref{eq:nn_g}) \\
$\alpha(E_{\rm{NN}})$ & Coefficient $\alpha$ in Eq.~(\ref{eq:nn_g}) \\
$\beta(E_{\rm{NN}})$ & Coefficient $\beta$ in Eq.~(\ref{eq:nn_g}) in fm$^{-3}$
\end{tabular*}
\label{tableI}

\renewcommand{\arraystretch}{1.0}

\bigskip

\section*{Tables 2--17.\label{tbl_cross} 
Elastic \raa\ scattering cross sections (normalized to
  the Rutherford cross section)}
\begin{tabular*}{0.95\textwidth}{@{}@{\extracolsep{\fill}}lp{5.5in}@{}}
$\vartheta_{\rm{c.m.}}$ & Scattering angle in the center-of-mass system in
  degrees \\
$\sigma/\sigma_R$ & Ratio of experimental differential cross section
  $\bigl(\frac{d\sigma}{d\Omega}\bigr)_{\rm{exp}}$ to the Rutherford cross
  section $\bigl(\frac{d\sigma}{d\Omega}\bigr)_R$ \\
$\Delta(\sigma/\sigma_R)$ & Uncertainty of the ratio $\sigma/\sigma_R$; the
  uncertainty of the scattering angle $\vartheta_{\rm{c.m.}}$ is included in
  $\Delta(\sigma/\sigma_R)$ using standard error propagation
\end{tabular*}
\label{tableII}

 \newpage

\GraphExplanation

\bigskip

\section*{Graphs 1--3. Elastic \raa\ scattering angular distributions for
  $^{89}$Y, $^{92}$Mo, $^{106,110,116}$Cd, $^{112,124}$Sn, and $^{144}$Sm,
  normalized to the Rutherford cross section.} 
The graphs show the experimental data of elastic \al\ scattering which are
also given in tabular form in Tables \ref{tab:y89e16} --
\ref{tab:sm144e20}. The angular distributions are normalized to Rutherford
scattering of point-like charges. The full lines result from a local fit with
the parameters given in Table \ref{tab:pot}. The dashed lines correspond to
the global potential in the Appendix (Sect.~\ref{sec:new}).

\datatables 



\setlength{\LTleft}{0pt}
\setlength{\LTright}{0pt} 


\setlength{\tabcolsep}{0.5\tabcolsep}

\renewcommand{\arraystretch}{1.0}

\footnotesize 

\begin{longtable}{@{\extracolsep\fill}rrlll}
\caption{Parameters of the energy dependence of the interaction. 
See page\ \pageref{tbl1te} for Explanation of Tables
\label{tab:interaction}}
\mbox{$E_{\rm{lab}}$ (MeV)} 
& \mbox{$E_{\rm{NN}}$ (MeV)} 
& \mbox{$C(E_{\rm{NN}})$}
& \mbox{$\alpha(E_{\rm{NN}})$}
& \mbox{$\beta(E_{\rm{NN}})$ (fm$^{-3}$)} \\
\hline\\
\endfirsthead\\
\hline\\
\endhead
   5.0 &    1.25 &   0.4347 &   4.3023 &  10.5479 \\
   6.0 &    1.50 &   0.4344 &   4.3008 &  10.5336 \\
   7.0 &    1.75 &   0.4341 &   4.2992 &  10.5193 \\
   8.0 &    2.00 &   0.4338 &   4.2977 &  10.5049 \\
   9.0 &    2.25 &   0.4335 &   4.2961 &  10.4904 \\
  10.0 &    2.50 &   0.4331 &   4.2944 &  10.4758 \\
  11.0 &    2.75 &   0.4328 &   4.2927 &  10.4612 \\
  12.0 &    3.00 &   0.4325 &   4.2910 &  10.4465 \\
\end{longtable}

\newpage

\renewcommand{\arraystretch}{1.0}

\begin{longtable}{@{\extracolsep\fill}rrrrrrrrr}
\caption{$^{89}$Y\raa $^{89}$Y elastic scattering cross section (normalized to
  the Rutherford cross section) at the energy $E = 15.51$\,MeV.
  See page\ \pageref{tbl_cross} for Explanation of Tables
\label{tab:y89e16}}
\mbox{$\vartheta_{\rm{c.m.}}$ (deg)} 
& \mbox{$\sigma/\sigma_R$} 
& \mbox{$\Delta(\sigma/\sigma_R)$}
& \mbox{$\vartheta_{\rm{c.m.}}$ (deg)} 
& \mbox{$\sigma/\sigma_R$} 
& \mbox{$\Delta(\sigma/\sigma_R)$} 
& \mbox{$\vartheta_{\rm{c.m.}}$ (deg)} 
& \mbox{$\sigma/\sigma_R$} 
& \mbox{$\Delta(\sigma/\sigma_R)$} \\
\hline\\
\endfirsthead\\
\hline\\
\endhead
  20.984 & 9.4408e-01 &  1.380e-02    &   82.557 & 3.5285e-01 &  5.820e-03  &  148.456 & 5.5380e-02 &  3.270e-03 \\
  21.991 & 9.2449e-01 &  1.350e-02    &   82.560 & 3.5216e-01 &  6.090e-03  &  149.790 & 5.3350e-02 &  3.080e-03 \\
  23.039 & 9.4977e-01 &  1.375e-02    &   83.561 & 3.3975e-01 &  5.570e-03  &  151.237 & 5.1430e-02 &  3.010e-03 \\
  24.040 & 9.1766e-01 &  1.339e-02    &   84.567 & 3.3009e-01 &  5.770e-03  &  153.311 & 6.0450e-02 &  2.670e-03 \\
  25.122 & 9.4139e-01 &  1.391e-02    &   85.571 & 3.0974e-01 &  5.090e-03  &  155.252 & 5.2120e-02 &  2.530e-03 \\
  26.158 & 9.3506e-01 &  1.387e-02    &   86.573 & 3.0324e-01 &  4.960e-03  &  157.151 & 5.3420e-02 &  2.400e-03 \\
  27.212 & 9.5078e-01 &  1.381e-02    &   87.575 & 2.9330e-01 &  4.710e-03  &  159.028 & 5.3050e-02 &  2.500e-03 \\
  28.223 & 9.4879e-01 &  1.392e-02    &   88.580 & 2.8679e-01 &  4.650e-03  &  160.879 & 5.7580e-02 &  2.710e-03 \\
\end{longtable}

\newpage

\renewcommand{\arraystretch}{1.0}

\begin{longtable}{@{\extracolsep\fill}rrrrrrrrr}
\caption{$^{89}$Y\raa $^{89}$Y elastic scattering cross section (normalized to
  the Rutherford cross section) at the energy $E = 18.63$\,MeV.
  See page\ \pageref{tbl_cross} for Explanation of Tables
\label{tab:y89e19}}
\mbox{$\vartheta_{\rm{c.m.}}$ (deg)} 
& \mbox{$\sigma/\sigma_R$} 
& \mbox{$\Delta(\sigma/\sigma_R)$}
& \mbox{$\vartheta_{\rm{c.m.}}$ (deg)} 
& \mbox{$\sigma/\sigma_R$} 
& \mbox{$\Delta(\sigma/\sigma_R)$} 
& \mbox{$\vartheta_{\rm{c.m.}}$ (deg)} 
& \mbox{$\sigma/\sigma_R$} 
& \mbox{$\Delta(\sigma/\sigma_R)$} \\
\hline\\
\endfirsthead\\
\hline\\
\endhead
  20.659 & 9.5125e-01 &  9.920e-03    &   82.491 & 1.2239e-01 &  2.120e-03  &  131.350 & 2.3220e-02 &  9.332e-04 \\
  21.766 & 9.9453e-01 &  1.072e-02    &   82.507 & 1.2153e-01 &  2.020e-03  &  132.820 & 2.0460e-02 &  9.068e-04 \\
  22.835 & 1.0425e+00 &  1.156e-02    &   83.517 & 1.1809e-01 &  2.480e-03  &  134.266 & 1.9290e-02 &  8.859e-04 \\
  23.838 & 1.0640e+00 &  1.287e-02    &   83.518 & 1.1550e-01 &  2.130e-03  &  135.737 & 1.9310e-02 &  8.613e-04 \\
  24.837 & 1.0802e+00 &  1.293e-02    &   84.527 & 1.1077e-01 &  1.890e-03  &  137.169 & 1.8130e-02 &  8.847e-04 \\
  25.911 & 1.0865e+00 &  1.320e-02    &   85.543 & 1.0711e-01 &  1.830e-03  &  138.120 & 1.7260e-02 &  7.674e-04 \\
  26.983 & 1.0968e+00 &  1.377e-02    &   86.546 & 1.0170e-01 &  1.750e-03  &  138.591 & 1.9420e-02 &  8.431e-04 \\
  27.992 & 1.0956e+00 &  1.382e-02    &   87.549 & 9.6730e-02 &  1.680e-03  &  139.524 & 1.9210e-02 &  8.144e-04 \\
\end{longtable}

\newpage

\renewcommand{\arraystretch}{1.0}

\begin{longtable}{@{\extracolsep\fill}rrrrrrrrr}
\caption{$^{92}$Mo\raa $^{92}$Mo elastic scattering cross section (normalized to
  the Rutherford cross section) at the energy $E = 13.20$\,MeV.
  See page\ \pageref{tbl_cross} for Explanation of Tables
\label{tab:mo92e13}}
\mbox{$\vartheta_{\rm{c.m.}}$ (deg)} 
& \mbox{$\sigma/\sigma_R$} 
& \mbox{$\Delta(\sigma/\sigma_R)$}
& \mbox{$\vartheta_{\rm{c.m.}}$ (deg)} 
& \mbox{$\sigma/\sigma_R$} 
& \mbox{$\Delta(\sigma/\sigma_R)$} 
& \mbox{$\vartheta_{\rm{c.m.}}$ (deg)} 
& \mbox{$\sigma/\sigma_R$} 
& \mbox{$\Delta(\sigma/\sigma_R)$} \\
\hline\\
\endfirsthead\\
\hline\\
\endhead
  20.145 & 9.4990e-01 &  2.168e-02    &   82.080 & 9.5869e-01 &  1.090e-02  &  141.615 & 4.3693e-01 &  5.190e-03 \\
  21.185 & 9.5014e-01 &  2.080e-02    &   82.081 & 9.5809e-01 &  1.073e-02  &  142.622 & 4.2179e-01 &  5.200e-03 \\
  22.226 & 9.5551e-01 &  2.011e-02    &   83.098 & 9.4920e-01 &  1.078e-02  &  144.065 & 4.1206e-01 &  5.310e-03 \\
  23.270 & 9.6942e-01 &  1.967e-02    &   84.107 & 9.3002e-01 &  1.006e-02  &  145.505 & 4.1225e-01 &  5.150e-03 \\
  24.311 & 9.6736e-01 &  1.897e-02    &   85.119 & 9.2954e-01 &  1.019e-02  &  146.947 & 4.0786e-01 &  5.170e-03 \\
  25.334 & 9.6642e-01 &  1.836e-02    &   86.132 & 9.1681e-01 &  9.890e-03  &  148.375 & 4.0670e-01 &  5.340e-03 \\
  26.374 & 9.7713e-01 &  1.801e-02    &   87.144 & 8.9780e-01 &  9.680e-03  &  149.798 & 4.0642e-01 &  5.090e-03 \\
  27.416 & 9.8018e-01 &  1.755e-02    &   88.154 & 8.9004e-01 &  9.610e-03  &  151.201 & 4.0148e-01 &  4.620e-03 \\
\end{longtable}

\newpage

\renewcommand{\arraystretch}{1.0}

\begin{longtable}{@{\extracolsep\fill}rrrrrrrrr}
\caption{$^{92}$Mo\raa $^{92}$Mo elastic scattering cross section (normalized to
  the Rutherford cross section) at the energy $E = 15.69$\,MeV.
  See page\ \pageref{tbl_cross} for Explanation of Tables
\label{tab:mo92e16}}
\mbox{$\vartheta_{\rm{c.m.}}$ (deg)} 
& \mbox{$\sigma/\sigma_R$} 
& \mbox{$\Delta(\sigma/\sigma_R)$}
& \mbox{$\vartheta_{\rm{c.m.}}$ (deg)} 
& \mbox{$\sigma/\sigma_R$} 
& \mbox{$\Delta(\sigma/\sigma_R)$} 
& \mbox{$\vartheta_{\rm{c.m.}}$ (deg)} 
& \mbox{$\sigma/\sigma_R$} 
& \mbox{$\Delta(\sigma/\sigma_R)$} \\
\hline\\
\endfirsthead\\
\hline\\
\endhead
  19.687 & 9.6174e-01 &  2.211e-02    &   80.972 & 5.4520e-01 &  6.100e-03  &  131.220 & 1.3984e-01 &  1.850e-03 \\
  20.730 & 9.7560e-01 &  2.148e-02    &   81.993 & 5.2115e-01 &  5.770e-03  &  132.663 & 1.3439e-01 &  1.930e-03 \\
  21.768 & 9.8174e-01 &  2.077e-02    &   81.994 & 5.2880e-01 &  5.990e-03  &  134.107 & 1.3244e-01 &  1.960e-03 \\
  22.817 & 9.7665e-01 &  1.989e-02    &   83.017 & 5.0550e-01 &  6.400e-03  &  135.549 & 1.3059e-01 &  1.920e-03 \\
  23.850 & 9.7427e-01 &  1.916e-02    &   84.039 & 4.9347e-01 &  5.500e-03  &  136.487 & 1.2878e-01 &  1.890e-03 \\
  24.903 & 9.5950e-01 &  1.825e-02    &   84.039 & 5.0465e-01 &  5.700e-03  &  136.984 & 1.2806e-01 &  1.850e-03 \\
  25.962 & 9.5684e-01 &  1.763e-02    &   85.060 & 4.7936e-01 &  5.340e-03  &  137.934 & 1.2528e-01 &  1.820e-03 \\
  26.998 & 9.4495e-01 &  1.691e-02    &   85.061 & 4.8557e-01 &  5.460e-03  &  138.421 & 1.2335e-01 &  1.780e-03 \\
\end{longtable}

\newpage

\renewcommand{\arraystretch}{1.0}

\begin{longtable}{@{\extracolsep\fill}rrrrrrrrr}
\caption{$^{92}$Mo\raa $^{92}$Mo elastic scattering cross section (normalized to
  the Rutherford cross section) at the energy $E = 18.70$\,MeV.
  See page\ \pageref{tbl_cross} for Explanation of Tables
\label{tab:mo92e19}}
\mbox{$\vartheta_{\rm{c.m.}}$ (deg)} 
& \mbox{$\sigma/\sigma_R$} 
& \mbox{$\Delta(\sigma/\sigma_R)$}
& \mbox{$\vartheta_{\rm{c.m.}}$ (deg)} 
& \mbox{$\sigma/\sigma_R$} 
& \mbox{$\Delta(\sigma/\sigma_R)$} 
& \mbox{$\vartheta_{\rm{c.m.}}$ (deg)} 
& \mbox{$\sigma/\sigma_R$} 
& \mbox{$\Delta(\sigma/\sigma_R)$} \\
\hline\\
\endfirsthead\\
\hline\\
\endhead
  20.388 & 9.2461e-01 &  2.174e-02    &   78.029 & 2.1927e-01 &  2.880e-03  &  140.274 & 3.0920e-02 &  9.450e-04 \\
  21.420 & 9.4039e-01 &  2.124e-02    &   79.034 & 2.1144e-01 &  2.800e-03  &  141.352 & 2.9660e-02 &  1.040e-03 \\
  22.467 & 9.5479e-01 &  2.077e-02    &   80.055 & 2.0495e-01 &  2.740e-03  &  141.682 & 3.1520e-02 &  7.390e-04 \\
  23.498 & 9.7596e-01 &  2.050e-02    &   81.070 & 1.9479e-01 &  2.630e-03  &  142.768 & 3.1070e-02 &  1.020e-03 \\
  24.561 & 1.0033e+00 &  2.037e-02    &   82.083 & 1.8778e-01 &  2.530e-03  &  144.268 & 2.8560e-02 &  9.600e-04 \\
  25.573 & 1.0211e+00 &  2.015e-02    &   82.086 & 1.9002e-01 &  2.570e-03  &  145.701 & 2.8850e-02 &  9.500e-04 \\
  26.634 & 1.0474e+00 &  2.009e-02    &   84.107 & 1.7900e-01 &  2.500e-03  &  147.150 & 3.3200e-02 &  1.090e-03 \\
  27.625 & 1.0512e+00 &  1.964e-02    &   85.127 & 1.7262e-01 &  2.480e-03  &  148.554 & 3.0880e-02 &  1.010e-03 \\
\end{longtable}

\newpage

\renewcommand{\arraystretch}{1.0}

\begin{longtable}{@{\extracolsep\fill}rrrrrrrrr}
\caption{$^{106}$Cd\raa $^{106}$Cd elastic scattering cross section
  (normalized to the Rutherford cross section) at the energy $E = 15.55$\,MeV.
  See page\ \pageref{tbl_cross} for Explanation of Tables
\label{tab:cd106e16}}
\mbox{$\vartheta_{\rm{c.m.}}$ (deg)} 
& \mbox{$\sigma/\sigma_R$} 
& \mbox{$\Delta(\sigma/\sigma_R)$}
& \mbox{$\vartheta_{\rm{c.m.}}$ (deg)} 
& \mbox{$\sigma/\sigma_R$} 
& \mbox{$\Delta(\sigma/\sigma_R)$} 
& \mbox{$\vartheta_{\rm{c.m.}}$ (deg)} 
& \mbox{$\sigma/\sigma_R$} 
& \mbox{$\Delta(\sigma/\sigma_R)$} \\
\hline\\
\endfirsthead\\
\hline\\
\endhead
  20.653 & 9.7300e-01 &  4.072e-02    &   86.837 & 7.5361e-01 &  1.276e-02  &  166.497 & 2.0154e-01 &  3.836e-03 \\
  21.684 & 9.8571e-01 &  3.955e-02    &   87.783 & 7.3586e-01 &  1.248e-02  &  168.427 & 1.9976e-01 &  4.162e-03 \\
  22.715 & 9.9891e-01 &  3.852e-02    &   88.643 & 7.2327e-01 &  1.213e-02  &  170.357 & 1.9959e-01 &  3.983e-03 \\
  23.746 & 9.9569e-01 &  3.698e-02    &   90.886 & 6.9154e-01 &  1.176e-02 \\
  24.777 & 9.9469e-01 &  3.569e-02    &   91.317 & 6.9152e-01 &  1.158e-02 \\
  25.807 & 9.9107e-01 &  3.434e-02    &   92.166 & 6.6226e-01 &  1.214e-02 \\
  26.837 & 9.9741e-01 &  3.348e-02    &   93.107 & 6.4761e-01 &  1.102e-02 \\
  27.867 & 1.0020e+00 &  3.264e-02    &   94.075 & 6.3464e-01 &  1.090e-02 \\
\end{longtable}

\newpage

\renewcommand{\arraystretch}{1.0}

\begin{longtable}{@{\extracolsep\fill}rrrrrrrrr}
\caption{$^{106}$Cd\raa $^{106}$Cd elastic scattering cross section
  (normalized to the Rutherford cross section) at the energy $E = 17.00$\,MeV.
  See page\ \pageref{tbl_cross} for Explanation of Tables
\label{tab:cd106e17}}
\mbox{$\vartheta_{\rm{c.m.}}$ (deg)} 
& \mbox{$\sigma/\sigma_R$} 
& \mbox{$\Delta(\sigma/\sigma_R)$}
& \mbox{$\vartheta_{\rm{c.m.}}$ (deg)} 
& \mbox{$\sigma/\sigma_R$} 
& \mbox{$\Delta(\sigma/\sigma_R)$} 
& \mbox{$\vartheta_{\rm{c.m.}}$ (deg)} 
& \mbox{$\sigma/\sigma_R$} 
& \mbox{$\Delta(\sigma/\sigma_R)$} \\
\hline\\
\endfirsthead\\
\hline\\
\endhead
  20.804 & 9.9992e-01 &  1.381e-02    &   82.140 & 5.4214e-01 &  8.042e-03  &  151.026 & 1.0101e-01 &  3.197e-03 \\
  21.832 & 1.0135e+00 &  1.394e-02    &   82.143 & 5.4672e-01 &  8.178e-03  &  151.064 & 8.9107e-02 &  2.051e-03 \\
  22.875 & 1.0163e+00 &  1.399e-02    &   83.145 & 5.2660e-01 &  7.758e-03  &  151.146 & 1.0069e-01 &  3.840e-03 \\
  23.891 & 1.0058e+00 &  1.384e-02    &   84.146 & 5.1823e-01 &  7.480e-03  &  153.058 & 9.8905e-02 &  3.644e-03 \\
  24.928 & 9.9453e-01 &  1.363e-02    &   85.153 & 4.9951e-01 &  7.368e-03  &  154.982 & 9.7807e-02 &  3.701e-03 \\
  25.956 & 9.8697e-01 &  1.345e-02    &   86.153 & 4.8068e-01 &  7.123e-03  &  156.900 & 9.5861e-02 &  5.296e-03 \\
  27.034 & 9.4819e-01 &  1.288e-02    &   87.159 & 4.5839e-01 &  6.791e-03  &  158.833 & 9.1021e-02 &  3.396e-03 \\
  28.029 & 9.5908e-01 &  1.301e-02    &   88.159 & 4.5435e-01 &  6.702e-03  &  160.679 & 8.8638e-02 &  3.115e-03 \\
\end{longtable}

\newpage

\renewcommand{\arraystretch}{1.0}

\begin{longtable}{@{\extracolsep\fill}rrrrrrrrr}
\caption{$^{106}$Cd\raa $^{106}$Cd elastic scattering cross section
  (normalized to the Rutherford cross section) at the energy $E = 18.88$\,MeV.
  See page\ \pageref{tbl_cross} for Explanation of Tables
\label{tab:cd106e19}}
\mbox{$\vartheta_{\rm{c.m.}}$ (deg)} 
& \mbox{$\sigma/\sigma_R$} 
& \mbox{$\Delta(\sigma/\sigma_R)$}
& \mbox{$\vartheta_{\rm{c.m.}}$ (deg)} 
& \mbox{$\sigma/\sigma_R$} 
& \mbox{$\Delta(\sigma/\sigma_R)$} 
& \mbox{$\vartheta_{\rm{c.m.}}$ (deg)} 
& \mbox{$\sigma/\sigma_R$} 
& \mbox{$\Delta(\sigma/\sigma_R)$} \\
\hline\\
\endfirsthead\\
\hline\\
\endhead
  20.715 & 9.5450e-01 &  4.471e-02    &   89.150 & 2.1613e-01 &  4.677e-03  &  170.303 & 2.9448e-02 &  8.698e-04 \\
  21.749 & 9.6545e-01 &  4.626e-02    &   90.115 & 2.1062e-01 &  4.447e-03 \\
  22.783 & 9.5007e-01 &  4.356e-02    &   91.077 & 1.9701e-01 &  4.139e-03 \\
  23.817 & 9.5505e-01 &  4.373e-02    &   92.076 & 1.9301e-01 &  4.274e-03 \\
  24.851 & 9.5502e-01 &  4.507e-02    &   93.076 & 1.8926e-01 &  4.505e-03 \\
  25.884 & 9.6959e-01 &  4.357e-02    &   94.076 & 1.8049e-01 &  4.247e-03 \\
  26.917 & 9.7426e-01 &  4.293e-02    &   95.075 & 1.7476e-01 &  4.064e-03 \\
  27.950 & 9.8297e-01 &  4.260e-02    &   96.074 & 1.6626e-01 &  3.800e-03 \\
\end{longtable}

\newpage

\renewcommand{\arraystretch}{1.0}

\begin{longtable}{@{\extracolsep\fill}rrrrrrrrr}
\caption{$^{110}$Cd\raa $^{110}$Cd elastic scattering cross section
  (normalized to the Rutherford cross section) at the energy $E = 15.56$\,MeV.
  See page\ \pageref{tbl_cross} for Explanation of Tables
\label{tab:cd110e16}}
\mbox{$\vartheta_{\rm{c.m.}}$ (deg)} 
& \mbox{$\sigma/\sigma_R$} 
& \mbox{$\Delta(\sigma/\sigma_R)$}
& \mbox{$\vartheta_{\rm{c.m.}}$ (deg)} 
& \mbox{$\sigma/\sigma_R$} 
& \mbox{$\Delta(\sigma/\sigma_R)$} 
& \mbox{$\vartheta_{\rm{c.m.}}$ (deg)} 
& \mbox{$\sigma/\sigma_R$} 
& \mbox{$\Delta(\sigma/\sigma_R)$} \\
\hline\\
\endfirsthead\\
\hline\\
\endhead
  20.799 & 9.9861e-01 &  4.249e-02    &  114.547 & 2.7345e-01 &  1.329e-02 \\
  21.832 & 1.0017e+00 &  4.262e-02    &  117.013 & 2.6117e-01 &  1.363e-02 \\
  22.865 & 1.0025e+00 &  4.266e-02    &  119.482 & 2.3777e-01 &  1.301e-02 \\
  23.898 & 1.0074e+00 &  4.287e-02    &  121.932 & 2.2044e-01 &  1.121e-02 \\
  24.930 & 1.0033e+00 &  4.271e-02    &  124.397 & 2.0393e-01 &  1.016e-02 \\
  25.962 & 1.0084e+00 &  4.295e-02    &  126.848 & 1.9562e-01 &  1.017e-02 \\
  26.994 & 1.0124e+00 &  4.312e-02    &  129.284 & 1.8326e-01 &  1.140e-02 \\
  28.026 & 1.0142e+00 &  4.322e-02    &  131.739 & 1.8652e-01 &  9.384e-03 \\
\end{longtable}

\newpage

\renewcommand{\arraystretch}{1.0}

\begin{longtable}{@{\extracolsep\fill}rrrrrrrrr}
\caption{$^{110}$Cd\raa $^{110}$Cd elastic scattering cross section
  (normalized to the Rutherford cross section) at the energy $E = 18.76$\,MeV.
  See page\ \pageref{tbl_cross} for Explanation of Tables
\label{tab:cd110e19}}
\mbox{$\vartheta_{\rm{c.m.}}$ (deg)} 
& \mbox{$\sigma/\sigma_R$} 
& \mbox{$\Delta(\sigma/\sigma_R)$}
& \mbox{$\vartheta_{\rm{c.m.}}$ (deg)} 
& \mbox{$\sigma/\sigma_R$} 
& \mbox{$\Delta(\sigma/\sigma_R)$} 
& \mbox{$\vartheta_{\rm{c.m.}}$ (deg)} 
& \mbox{$\sigma/\sigma_R$} 
& \mbox{$\Delta(\sigma/\sigma_R)$} \\
\hline\\
\endfirsthead\\
\hline\\
\endhead
  20.799 & 9.9928e-01 &  4.258e-02    &   92.188 & 1.7828e-01 &  8.803e-03 \\
  21.832 & 9.9744e-01 &  4.252e-02    &   93.182 & 1.7726e-01 &  8.590e-03 \\
  22.865 & 9.8590e-01 &  4.204e-02    &   94.182 & 1.6466e-01 &  8.288e-03 \\
  23.898 & 9.8193e-01 &  4.190e-02    &   96.180 & 1.4559e-01 &  8.127e-03 \\
  24.930 & 9.6942e-01 &  4.141e-02    &   98.176 & 1.4387e-01 &  7.888e-03 \\
  25.962 & 9.7365e-01 &  4.158e-02    &  100.170 & 1.2732e-01 &  7.227e-03 \\
  26.994 & 9.7199e-01 &  4.154e-02    &  102.161 & 1.1692e-01 &  6.611e-03 \\
  28.026 & 9.8868e-01 &  4.230e-02    &  104.657 & 1.0127e-01 &  5.634e-03 \\
\end{longtable}

\newpage

\renewcommand{\arraystretch}{1.0}

\begin{longtable}{@{\extracolsep\fill}rrrrrrrrr}
\caption{$^{116}$Cd\raa $^{116}$Cd elastic scattering cross section
  (normalized to the Rutherford cross section) at the energy $E = 15.59$\,MeV.
  See page\ \pageref{tbl_cross} for Explanation of Tables
\label{tab:cd116e16}}
\mbox{$\vartheta_{\rm{c.m.}}$ (deg)} 
& \mbox{$\sigma/\sigma_R$} 
& \mbox{$\Delta(\sigma/\sigma_R)$}
& \mbox{$\vartheta_{\rm{c.m.}}$ (deg)} 
& \mbox{$\sigma/\sigma_R$} 
& \mbox{$\Delta(\sigma/\sigma_R)$} 
& \mbox{$\vartheta_{\rm{c.m.}}$ (deg)} 
& \mbox{$\sigma/\sigma_R$} 
& \mbox{$\Delta(\sigma/\sigma_R)$} \\
\hline\\
\endfirsthead\\
\hline\\
\endhead
  20.764 & 1.0064e+00 &  4.286e-02    &   87.068 & 6.1461e-01 &  2.717e-02 \\
  21.796 & 9.9886e-01 &  4.255e-02    &   88.071 & 6.0103e-01 &  2.660e-02 \\
  22.827 & 9.9380e-01 &  4.234e-02    &   89.071 & 5.7731e-01 &  2.559e-02 \\
  23.858 & 9.9514e-01 &  4.241e-02    &   89.874 & 5.5913e-01 &  2.472e-02 \\
  24.889 & 1.0040e+00 &  4.280e-02    &   91.086 & 5.5104e-01 &  2.435e-02 \\
  25.919 & 1.0062e+00 &  4.290e-02    &   92.080 & 5.3365e-01 &  2.392e-02 \\
  26.950 & 1.0054e+00 &  4.288e-02    &   93.074 & 5.1455e-01 &  2.317e-02 \\
  27.980 & 1.0109e+00 &  4.311e-02    &   94.074 & 5.0988e-01 &  2.309e-02 \\
\end{longtable}

\newpage

\renewcommand{\arraystretch}{1.0}

\begin{longtable}{@{\extracolsep\fill}rrrrrrrrr}
\caption{$^{116}$Cd\raa $^{116}$Cd elastic scattering cross section
  (normalized to the Rutherford cross section) at the energy $E = 18.80$\,MeV.
  See page\ \pageref{tbl_cross} for Explanation of Tables
\label{tab:cd116e19}}
\mbox{$\vartheta_{\rm{c.m.}}$ (deg)} 
& \mbox{$\sigma/\sigma_R$} 
& \mbox{$\Delta(\sigma/\sigma_R)$}
& \mbox{$\vartheta_{\rm{c.m.}}$ (deg)} 
& \mbox{$\sigma/\sigma_R$} 
& \mbox{$\Delta(\sigma/\sigma_R)$} 
& \mbox{$\vartheta_{\rm{c.m.}}$ (deg)} 
& \mbox{$\sigma/\sigma_R$} 
& \mbox{$\Delta(\sigma/\sigma_R)$} \\
\hline\\
\endfirsthead\\
\hline\\
\endhead
  20.764 & 1.0072e+00 &  4.292e-02    &   87.068 & 1.9946e-01 &  9.432e-03 \\
  21.796 & 9.9159e-01 &  4.226e-02    &   88.071 & 1.8615e-01 &  9.050e-03 \\
  22.827 & 9.8241e-01 &  4.188e-02    &   89.071 & 1.7876e-01 &  8.759e-03 \\
  23.858 & 9.6449e-01 &  4.116e-02    &   89.874 & 1.7302e-01 &  8.249e-03 \\
  24.889 & 9.7731e-01 &  4.172e-02    &   91.086 & 1.6651e-01 &  7.875e-03 \\
  25.919 & 1.0084e+00 &  4.305e-02    &   92.080 & 1.5707e-01 &  7.816e-03 \\
  26.950 & 1.0085e+00 &  4.308e-02    &   93.074 & 1.4978e-01 &  7.355e-03 \\
  27.980 & 1.0294e+00 &  4.398e-02    &   94.074 & 1.4642e-01 &  7.164e-03 \\
\end{longtable}

\newpage

\renewcommand{\arraystretch}{1.0}

\begin{longtable}{@{\extracolsep\fill}rrrrrrrrr}
\caption{$^{112}$Sn\raa $^{112}$Sn elastic scattering cross section
  (normalized to the Rutherford cross section) at the energy $E = 13.90$\,MeV.
  See page\ \pageref{tbl_cross} for Explanation of Tables
\label{tab:sn112e14}}
\mbox{$\vartheta_{\rm{c.m.}}$ (deg)} 
& \mbox{$\sigma/\sigma_R$} 
& \mbox{$\Delta(\sigma/\sigma_R)$}
& \mbox{$\vartheta_{\rm{c.m.}}$ (deg)} 
& \mbox{$\sigma/\sigma_R$} 
& \mbox{$\Delta(\sigma/\sigma_R)$} 
& \mbox{$\vartheta_{\rm{c.m.}}$ (deg)} 
& \mbox{$\sigma/\sigma_R$} 
& \mbox{$\Delta(\sigma/\sigma_R)$} \\
\hline\\
\endfirsthead\\
\caption[]{(continued)}
\mbox{$\vartheta_{\rm{c.m.}}$ (deg)} 
& \mbox{$\sigma/\sigma_R$} 
& \mbox{$\Delta(\sigma/\sigma_R)$}
& \mbox{$\vartheta_{\rm{c.m.}}$ (deg)} 
& \mbox{$\sigma/\sigma_R$} 
& \mbox{$\Delta(\sigma/\sigma_R)$} 
& \mbox{$\vartheta_{\rm{c.m.}}$ (deg)} 
& \mbox{$\sigma/\sigma_R$} 
& \mbox{$\Delta(\sigma/\sigma_R)$} \\
\hline\\
\endhead
  20.892 & 1.0000e+00 &  1.563e-02    &   82.050 & 9.9177e-01 &  1.565e-02  &  132.399 & 7.0151e-01 &  1.083e-02 \\
  21.892 & 9.8908e-01 &  1.545e-02    &   82.052 & 1.0025e+00 &  1.398e-02  &  133.396 & 7.1576e-01 &  1.108e-02 \\
  22.921 & 9.9991e-01 &  1.559e-02    &   83.054 & 1.0096e+00 &  1.597e-02  &  134.334 & 7.0760e-01 &  1.106e-02 \\
  23.984 & 1.0025e+00 &  1.561e-02    &   84.055 & 9.7749e-01 &  1.547e-02  &  134.572 & 7.1015e-01 &  1.424e-02 \\
  24.994 & 9.8886e-01 &  1.543e-02    &   85.067 & 1.0045e+00 &  1.578e-02  &  135.293 & 6.9623e-01 &  1.089e-02 \\
  26.058 & 1.0034e+00 &  1.683e-02    &   86.074 & 9.9044e-01 &  1.575e-02  &  135.594 & 7.0221e-01 &  1.478e-02 \\
  27.045 & 9.9602e-01 &  1.556e-02    &   87.069 & 9.9833e-01 &  1.587e-02  &  136.283 & 7.0071e-01 &  1.066e-02 \\
  28.086 & 9.9342e-01 &  1.552e-02    &   88.062 & 9.7902e-01 &  1.555e-02  &  136.616 & 6.9411e-01 &  1.344e-02 \\
\end{longtable}

\newpage

\renewcommand{\arraystretch}{1.0}

\begin{longtable}{@{\extracolsep\fill}rrrrrrrrr}
\caption{$^{112}$Sn\raa $^{112}$Sn elastic scattering cross section
  (normalized to the Rutherford cross section) at the energy $E = 18.84$\,MeV.
  See page\ \pageref{tbl_cross} for Explanation of Tables
\label{tab:sn112e19}}
\mbox{$\vartheta_{\rm{c.m.}}$ (deg)} 
& \mbox{$\sigma/\sigma_R$} 
& \mbox{$\Delta(\sigma/\sigma_R)$}
& \mbox{$\vartheta_{\rm{c.m.}}$ (deg)} 
& \mbox{$\sigma/\sigma_R$} 
& \mbox{$\Delta(\sigma/\sigma_R)$} 
& \mbox{$\vartheta_{\rm{c.m.}}$ (deg)} 
& \mbox{$\sigma/\sigma_R$} 
& \mbox{$\Delta(\sigma/\sigma_R)$} \\
\hline\\
\endfirsthead\\
\hline\\
\endhead
  20.837 & 1.0084e+00 &  1.457e-02    &   80.969 & 3.7159e-01 &  5.780e-03  &  141.404 & 5.5760e-02 &  1.180e-03 \\
  21.834 & 1.0018e+00 &  1.440e-02    &   81.030 & 3.7843e-01 &  5.420e-03  &  142.382 & 5.5590e-02 &  1.440e-03 \\
  22.872 & 9.9926e-01 &  1.435e-02    &   81.981 & 3.5955e-01 &  5.640e-03  &  143.859 & 5.2830e-02 &  1.380e-03 \\
  23.901 & 9.7235e-01 &  1.393e-02    &   82.031 & 3.6386e-01 &  5.320e-03  &  145.311 & 5.3560e-02 &  1.380e-03 \\
  24.937 & 9.9452e-01 &  1.422e-02    &   82.040 & 3.7327e-01 &  5.520e-03  &  146.757 & 5.1960e-02 &  1.350e-03 \\
  25.913 & 9.7006e-01 &  1.383e-02    &   82.056 & 3.7003e-01 &  5.630e-03  &  148.212 & 4.9170e-02 &  1.310e-03 \\
  26.983 & 9.5455e-01 &  1.358e-02    &   82.057 & 3.6668e-01 &  5.840e-03  &  149.660 & 5.0380e-02 &  1.360e-03 \\
  28.015 & 9.5034e-01 &  1.355e-02    &   83.041 & 3.5371e-01 &  5.290e-03  &  150.934 & 4.6900e-02 &  1.030e-03 \\
\end{longtable}

\newpage

\renewcommand{\arraystretch}{1.0}

\begin{longtable}{@{\extracolsep\fill}rrrrrrrrr}
\caption{$^{124}$Sn\raa $^{124}$Sn elastic scattering cross section
  (normalized to the Rutherford cross section) at the energy $E = 18.90$\,MeV.
  See page\ \pageref{tbl_cross} for Explanation of Tables
\label{tab:sn124e19}}
\mbox{$\vartheta_{\rm{c.m.}}$ (deg)} 
& \mbox{$\sigma/\sigma_R$} 
& \mbox{$\Delta(\sigma/\sigma_R)$}
& \mbox{$\vartheta_{\rm{c.m.}}$ (deg)} 
& \mbox{$\sigma/\sigma_R$} 
& \mbox{$\Delta(\sigma/\sigma_R)$} 
& \mbox{$\vartheta_{\rm{c.m.}}$ (deg)} 
& \mbox{$\sigma/\sigma_R$} 
& \mbox{$\Delta(\sigma/\sigma_R)$} \\
\hline\\
\endfirsthead\\
\hline\\
\endhead
  20.755 & 9.9945e-01 &  1.400e-02    &   77.834 & 3.7342e-01 &  5.390e-03  &  143.739 & 3.9730e-02 &  1.370e-03 \\
  21.806 & 9.9851e-01 &  1.395e-02    &   78.837 & 3.5055e-01 &  5.060e-03  &  145.193 & 3.7720e-02 &  1.310e-03 \\
  22.781 & 9.6804e-01 &  1.350e-02    &   79.844 & 3.3508e-01 &  4.780e-03  &  146.657 & 3.7890e-02 &  1.340e-03 \\
  23.877 & 9.9293e-01 &  1.382e-02    &   80.843 & 3.2126e-01 &  4.620e-03  &  148.113 & 3.4900e-02 &  1.250e-03 \\
  23.913 & 9.9293e-01 &  1.384e-02    &   81.844 & 3.0653e-01 &  4.490e-03  &  149.580 & 3.5640e-02 &  1.240e-03 \\
  24.854 & 9.4869e-01 &  1.319e-02    &   81.845 & 3.0476e-01 &  4.480e-03  &  151.037 & 3.5070e-02 &  1.240e-03 \\
  24.904 & 9.6870e-01 &  1.346e-02    &   82.846 & 2.9317e-01 &  4.320e-03  &  152.983 & 3.2820e-02 &  7.305e-04 \\
  25.874 & 9.6135e-01 &  1.333e-02    &   83.851 & 2.8308e-01 &  4.190e-03  &  154.920 & 3.2370e-02 &  6.963e-04 \\
\end{longtable}

\newpage

\renewcommand{\arraystretch}{1.0}

\begin{longtable}{@{\extracolsep\fill}rrrrrrrrr}
\caption{$^{144}$Sm\raa $^{144}$Sm elastic scattering cross section
  (normalized to the Rutherford cross section) at the energy $E = 19.45$\,MeV.
  See page\ \pageref{tbl_cross} for Explanation of Tables
\label{tab:sm144e20}}
\mbox{$\vartheta_{\rm{c.m.}}$ (deg)} 
& \mbox{$\sigma/\sigma_R$} 
& \mbox{$\Delta(\sigma/\sigma_R)$}
& \mbox{$\vartheta_{\rm{c.m.}}$ (deg)} 
& \mbox{$\sigma/\sigma_R$} 
& \mbox{$\Delta(\sigma/\sigma_R)$} 
& \mbox{$\vartheta_{\rm{c.m.}}$ (deg)} 
& \mbox{$\sigma/\sigma_R$} 
& \mbox{$\Delta(\sigma/\sigma_R)$} \\
\hline\\
\endfirsthead\\
\hline\\
\endhead
  15.023 & 9.9443e-01 &  1.136e-02    &   81.188 & 8.2501e-01 &  1.129e-02  &  149.154 & 1.5125e-01 &  1.919e-03 \\
  16.050 & 1.0003e+00 &  1.137e-02    &   82.193 & 8.0272e-01 &  8.556e-03  &  151.106 & 1.4843e-01 &  2.374e-03 \\
  17.076 & 9.9475e-01 &  1.145e-02    &   83.197 & 7.8205e-01 &  8.455e-03  &  153.057 & 1.4558e-01 &  1.767e-03 \\
  18.103 & 9.9632e-01 &  1.148e-02    &   84.201 & 7.7795e-01 &  8.465e-03  &  155.007 & 1.4175e-01 &  1.525e-03 \\
  19.129 & 1.0040e+00 &  1.137e-02    &   85.204 & 7.5133e-01 &  8.211e-03  &  156.957 & 1.4222e-01 &  1.730e-03 \\
  20.156 & 1.0128e+00 &  1.100e-02    &   86.207 & 7.3524e-01 &  7.913e-03  &  158.905 & 1.3723e-01 &  1.522e-03 \\
  21.182 & 1.0036e+00 &  1.037e-02    &   87.210 & 7.1104e-01 &  7.781e-03  &  160.853 & 1.3005e-01 &  1.662e-03 \\
  22.208 & 9.9883e-01 &  1.032e-02    &   88.211 & 6.9320e-01 &  7.676e-03  &  162.123 & 1.2510e-01 &  1.947e-03 \\
\end{longtable}

\newpage

\begin{Dfigures}[ht!]
\centering
\includegraphics{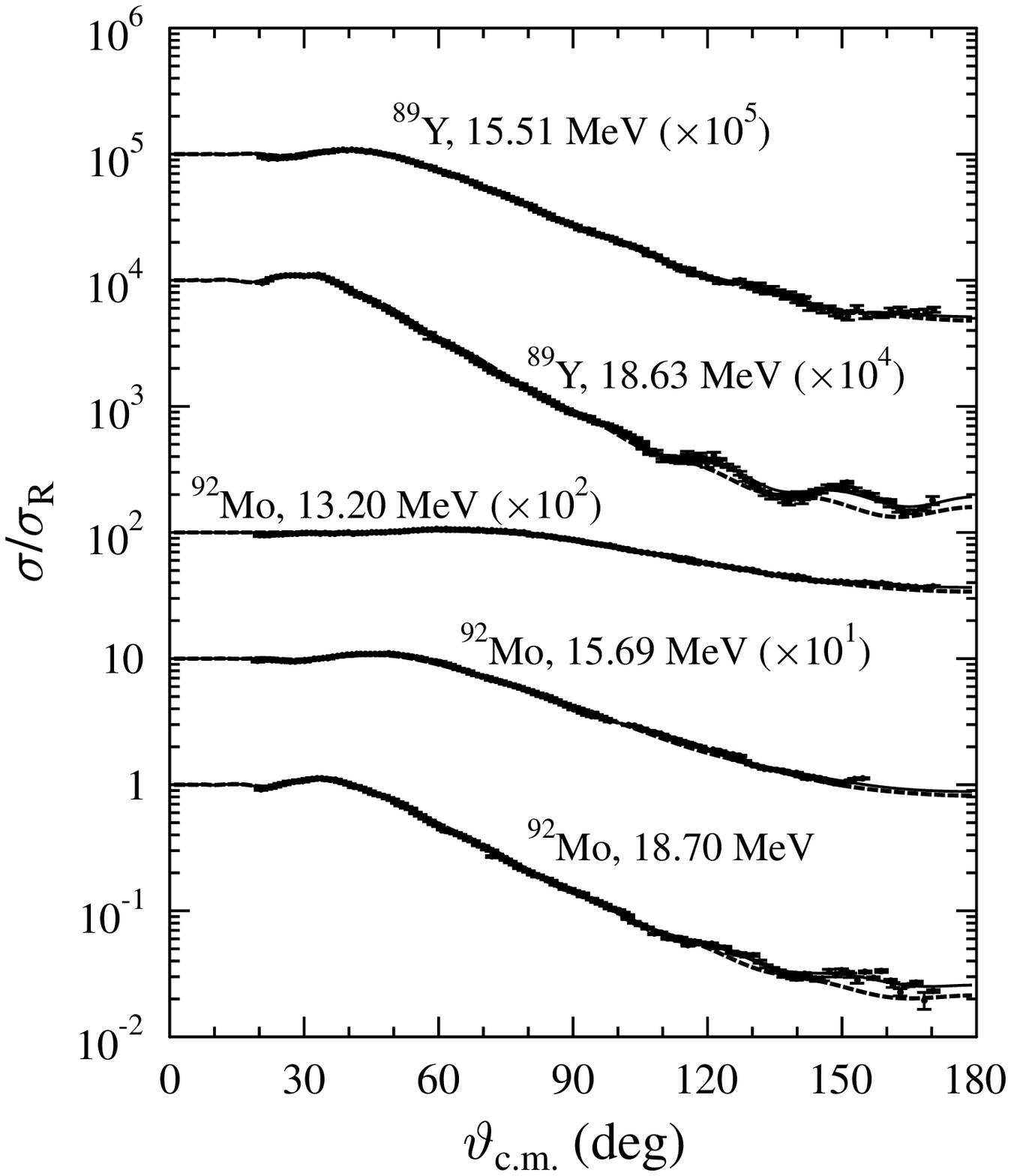}
\caption{Elastic \raa\ scattering angular distributions of $^{89}$Y and
  $^{92}$Mo. 
}
\label{fig:y89mo92}
\end{Dfigures}

\newpage

\begin{Dfigures}[ht!]
\centering
\includegraphics{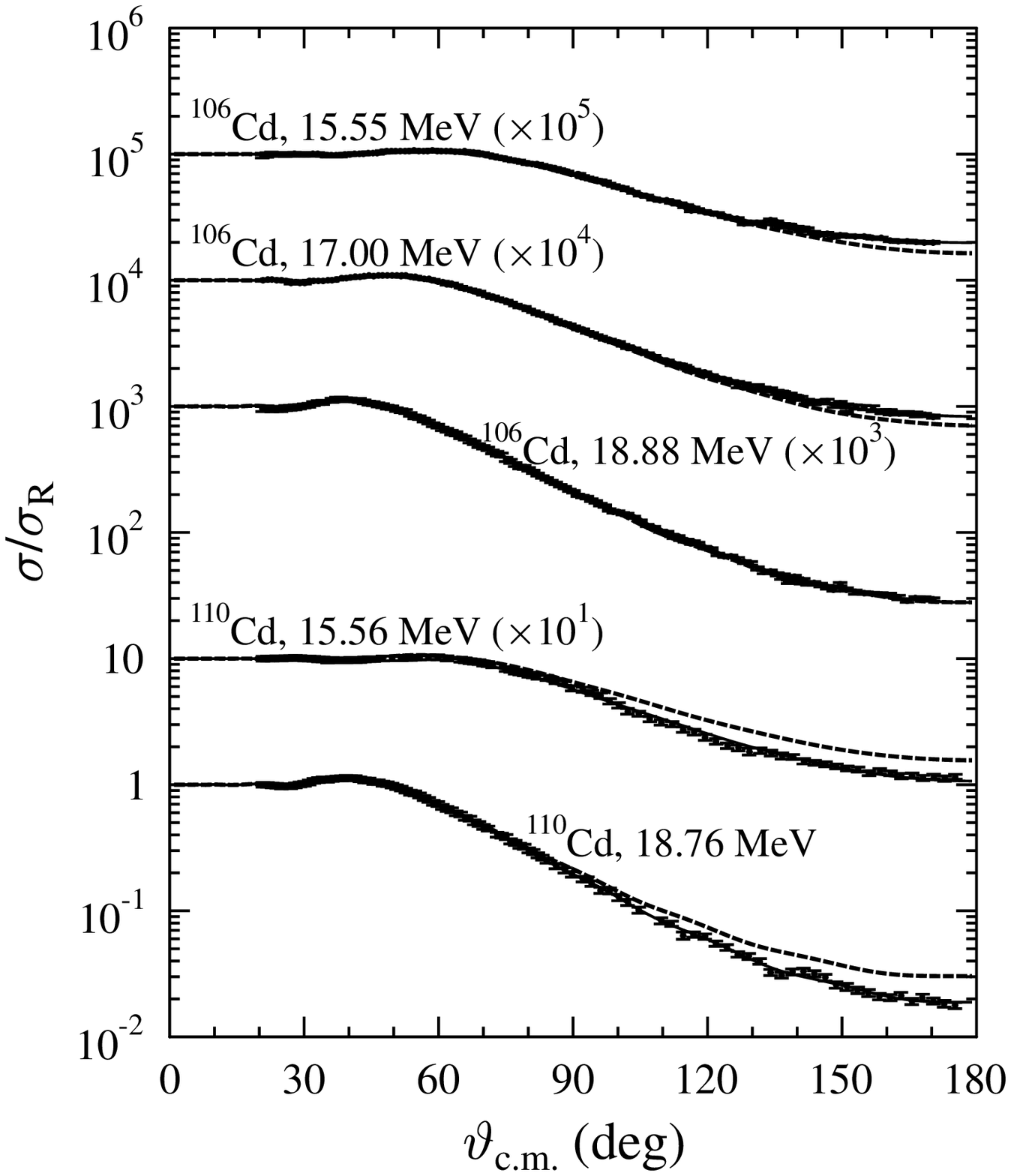}
\caption{Elastic \raa\ scattering angular distributions of $^{106}$Cd and
  $^{110}$Cd. 
}
\label{fig:cd106cd110}
\end{Dfigures}

\newpage

\begin{Dfigures}[ht!]
\centering
\includegraphics{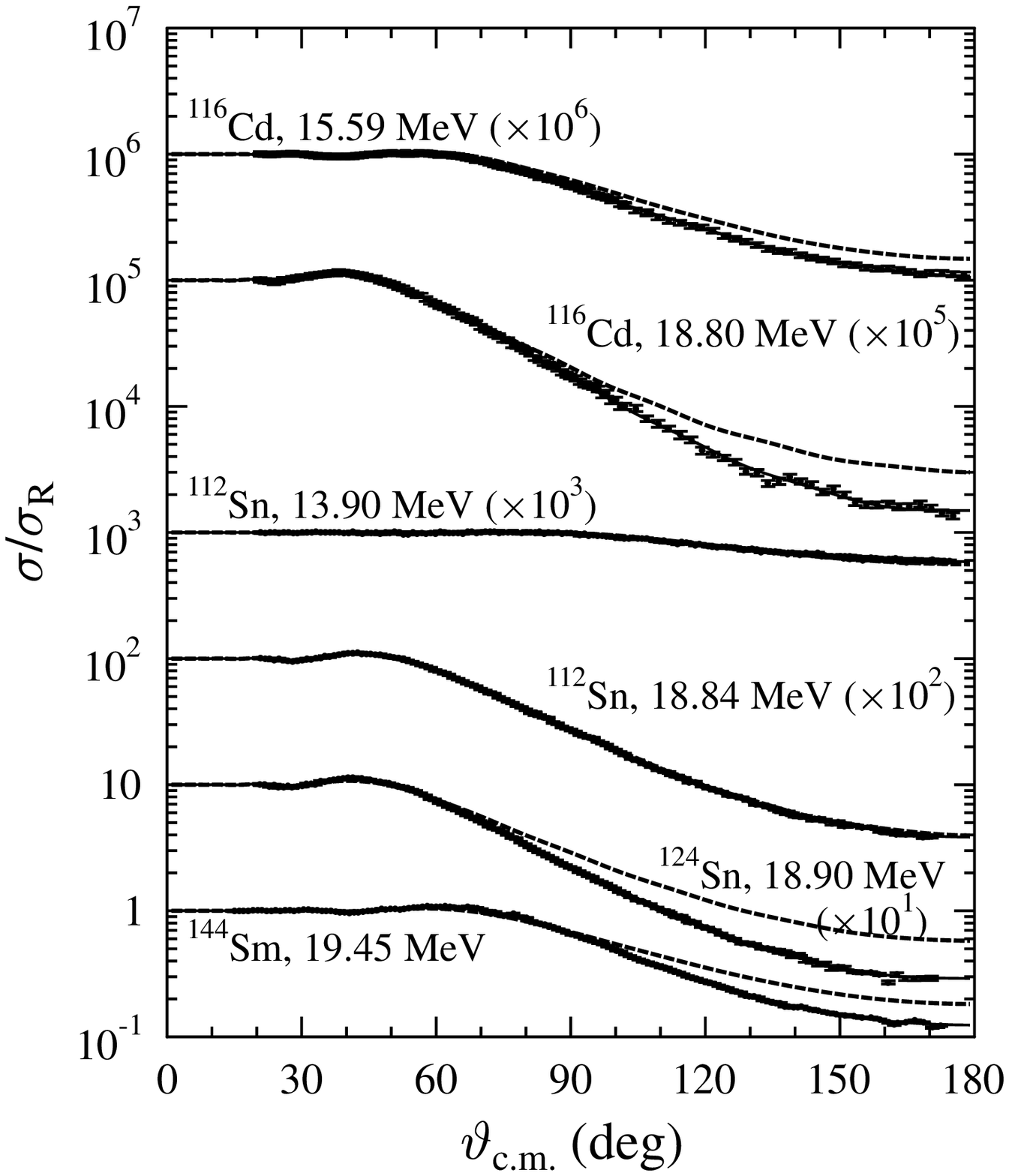}
\caption{Elastic \raa\ scattering angular distributions of $^{116}$Cd,
  $^{112}$Sn, $^{124}$Sn, and $^{144}$Sm.
}
\label{fig:cd116sm144}
\end{Dfigures}

\end{document}